\documentclass[structabstract]{aa}  
\usepackage{graphicx}
\usepackage{txfonts}

\usepackage{natbib}
\begin{document}
\title{The Planetary Nebula Spectrograph survey of S0 galaxy kinematics}

   \subtitle{Data and Overview}

   \author{A.~Cortesi\inst{1,3,2},   M.~Arnaboldi\inst{1}, L.~Coccato\inst{1}, M.R.~Merrifield\inst{2}, O.~Gerhard\inst{3}, 
   S.~Bamford\inst{2}, A.~J.~Romanowsky\inst{4,5}, N.~R.~Napolitano\inst{6}, N.~G.~Douglas\inst{7}, 
    K.~Kuijken\inst{8},  M.~Capaccioli\inst{9}, K.~C.~Freeman\inst{10},  A.~L.~Chies-Santos\inst{2} and V. Pota\inst{11}
                    }

              \institute{ European Southern Observatory, Karl-Schwarzschild-Strasse 2, 85748 Garching, Germany \\
             \email{aricorte@gmail.com}
             \and
             University of Nottingham, School of Physics and Astronomy, University Park, NG7 2RD Nottingham, UK \\ 
             \and
              Max-Planck-Institut f\"ur Extraterrestrische Physik,
              Giessenbachstrasse,  85741 Garching, Germany \\
              \and
              University of California Observatories, 1156 High Street, Santa Cruz, CA 95064, USA \\
              \and
              Department of Physics and Astronomy, San Jos\'e State University, One Washington Square, San Jose, CA 95192, USA \\
              \and
              Istituto Nazionale di Astrofisica, Osservatorio Astronomico di
              Capodimonte, Via Moiariello 16, 80131 Naples, Italy \\
              \and
              Kapteyn Astronomical Institute, University of Groningen, PO Box 800, 9700 AV Groningen, The Netherlands \\
              \and
              Leiden Observatory, Leiden University, PO Box 9513, 2300 RA Leiden, The Netherlands \\
              \and
               Dipartimento di Fisica, Universit\`a ``Federico II'', Naples, Italy \\
              \and
              Research School of Astronomy and Astrophysics, Australian National
              University, Canberra, Australia\\
              \and
              Centre for Astrophysics \& Supercomputing, Swinburne University, Hawthorn VIC 3122, Australia\\
             }

   \date{Received 29 August 2012 / Accepted 10 October 2012}

 
  \abstract
  {The origins of S0 galaxies remain obscure, with various mechanisms
    proposed for their formation, likely depending on environment.
    These mechanisms would imprint different signatures in the
    galaxies' stellar kinematics out to large radii, offering a method
    for distinguishing between them.}
  {We aim to study a sample of six S0 galaxies from a range of
    environments, and use planetary nebulae (PNe) as tracers of their
    stellar populations out to very large radii, to determine their
    kinematics in order to understand their origins.}
      {Using a special-purpose instrument, the Planetary Nebula
        Spectrograph, we observe and extract PNe catalogues for
        these six systems \thanks{(Tables 3 $\div$ 7 are only available in electronic form
at the CDS via anonymous ftp to cdsarc.u-strasbg.fr (130.79.128.5)
or via http://cdsweb.u-strasbg.fr/cgi-bin/qcat?J/A+A/) }.}
      { We show that the  PNe  have the same spatial distribution
        as the starlight, that the numbers of them are consistent with
        what would be expected in a comparable old stellar population
        in elliptical galaxies, and that their kinematics join
        smoothly onto those derived at smaller radii from conventional
        spectroscopy. }
    { The high-quality kinematic observations presented here form
        an excellent set for studying the detailed kinematics of S0
        galaxies, in order to unravel their formation histories. We
        find that PNe are good tracers of stellar kinematics in these
        systems. We show that the recovered kinematics are largely
        dominated by rotational motion, although with significant
        random velocities in most cases. }


   \titlerunning{PN.S survey of S0 galaxies}
   \authorrunning{Cortesi et al.}

   \maketitle
%

\section{Introduction}
\label{sec:introduction}

S0 or lenticular galaxies seem to occupy a rather important place in
the pantheon of galaxy types.  Their smooth appearance coupled
with their disk morphologies led \citet{Hubble36} to place them at the
centre of his sequence of galaxies between the smooth ellipticals and
the disk-dominated spirals, suggesting that they may provide a key
clue to understanding the relationship between these galaxies.
Further, \citet{Dressler80} established that in all dense environments
S0 galaxies are the most numerous luminous systems, outnumbering even
the ellipticals in the centres of clusters, so clearly any theory of
galaxy evolution has to contain a convincing explanation for the
origins of S0 galaxies.  

It is therefore rather frustrating that there is no consensus as to
how these systems form.  The lack of recent star formations implies
that they have not had ready access to a supply of cold gas in their recent history,
 but the reason for this lack of raw material remains a matter of debate.
The quenching of star formation could be the result of galaxy
minor mergers \citep{Bournaud}, analogous to the manner that more
major mergers are believed to shut down star formation in elliptical
galaxies.  Alternatively, a normal spiral galaxy could simply have had
its gas stripped out by any of a range of more gentle processes such
as ram-pressure stripping or tidal interactions
\citep{Gunn,Salamanca,Quilis,Kronberger,Byrd}.  Thus, the most basic
question of whether S0 galaxies are more closely related to
ellipticals or spirals remains unanswered, as does the issue of
whether multiple routes might lead to their formation depending on
their surroundings.

Fortunately, there are as-yet largely untapped clues to S0 formation
contained in their stellar kinematics.  The simpler gas stripping
processes would not have much impact on the collisionless stellar
component, so one would expect an S0 formed in this way to have the
same stellar kinematics as its progenitor spiral \citep{Salamanca,
  Williams}. On the other hand, a minor merger will heat the kinematics of the stellar
component as well, leading to disks with substantially more random
motion than is found in spirals \citep{Bournaud}.  More generally,
elliptical and spiral galaxies obey quite tight scaling relations that
relate their kinematics to their photometric properties, so by
investigating where S0s fall on these relations we can obtain a simple
model-independent insight as to how closely they are related to the
other galaxy types, and whether those connections vary with
environment.

The main problem in exploiting this information is that it is
observationally quite challenging to extract.  The outer regions of S0 disks
 are very faint, so conventional absorption-line spectroscopy
struggles to obtain data of sufficient depth to derive reliable disk
kinematics.  Further, the composite nature of these systems, with
varying contributions to the kinematics from disk and bulge at all
radii, means that very high quality spectral data would be required to
separate out the components reliably.

Fortunately, planetary nebulae (PNe) offer a simple alternative tracer
of stellar kinematics out to very large radii.  The strong emission
from the  [\ion{O}{iii}] 5007\,\AA\ line in PNe makes them easy to
identify even at very low surface brightnesses, and also to measure
their line-of-sight velocities from the Doppler shift in the line.  To
carry out both identification and kinematic measurement in a single
efficient step, we designed a special-purpose instrument, the
Planetary Nebula Spectrograph [PN.S; \citet{Nigel}], as a visitor
instrument for the 4.2m William Herschel Telescope (WHT) of the Isaac
Newton Group in La Palma.  Although originally intended to study
elliptical galaxies
\citep{Romanowsky,Nigel2,Coccato,Napolitano2,Napolitano3}, the PN.S has
already proved to be very effective in exploring the disks of spiral
galaxies \citep{Merrett}, so extending its work to S0s was an obvious
next step.

  As a pilot study to test this possibility, we observed the
  archetypal S0 galaxy NGC~1023 \citep{Ari}, and ascertained that it
  is possible to separate the more complex composite kinematics in
  these systems using a maximum likelihood method, which was shown
  to give more reliable results than a conventional analysis
  \citep{EDO}.  Based on the success of this proof-of-concept, we
  decided to carry out a more extensive PN.S study of 6 S0 galaxies
  that reside in a variety of environments, to see if this factor
  drives any variations between them, and have obtained observations
  with PN.S of each galaxy of sufficient depth to obtain samples of
  typically 100 PNe in each.

  In this paper we present the
  data from the survey and assess its reliability as a kinematic tracer in all such
  systems. Indeed, one concern raised with using PNe to measure
  stellar kinematics is that our understanding of the progenitors of
  these objects is somewhat incomplete, so it is possible that they do
  not form a representative sample of the underlying stellar
  population \citep{Dekel+05}.  This concern has been largely
  addressed for elliptical systems, where the PNe follow the same
  density profile and kinematics as the old stellar population where
  they overlap, and have consistent specific frequencies implying that
  they are not some random scattering of younger stars
  \citep{Coccato}.  However, S0s are a different class of system, so
  this issue needs to be looked at for these objects as well, to check
  that the concern can be similarly dismissed. The PN data sets are also more broadly useful to the galaxy formation community for studying the outer regions of S0s \citep[e.g.][]{Forbes2012}.  

  Accordingly, this paper presents a description of the sample itself
  (Section~\ref{sec:sample}) and the resulting catalogues
  (Section~\ref{sec:catalogues}).  We also carry out the necessary
  checks on the viability of PNe as tracers of the stellar population
  by comparing their spatial distribution to that of the starlight,
  and, quantifying the normalisation of this comparison, the specific
  frequency of PNe (Section~\ref{sec:PNe surface density}).  We
  present the resulting basic kinematics of the galaxies here
  (Section~\ref{sec:velocity profile}) and draw some initial
  conclusions (Section~\ref{sec:conclusion}). In the interests
  of clarity and the timeliness of the data, we defer the detailed
  modelling of the data for the full sample, together with its
  interpretation in terms of our understanding of S0 galaxy formation,
  to a subsequent paper.  

\section{The S0 sample}
\label{sec:sample}
This project aims to study a sample of S0 galaxies that
spans a range of environments, to investigate whether their properties vary
systematically with their surroundings, as would be expected if there are
multiple environmentally-driven channels for S0 formation.  As usual
in astronomy, we are faced by various conflicting pressures in
selecting such a sample.  First, in order to obtain reasonable numbers
of PNe, the galaxies have to be fairly nearby, but this means that we
cannot probe the rare densest environments of rich clusters.
Accordingly, we compromise by studying a sample that spans the widest
range available to us, from isolated S0s to poor groups to rich
groups.  Second, the requisite observations take a long time even with
an instrument as efficient as PN.S, so the sample has to be quite
modest in size, but there is also likely to be intrinsic variance due
to the personal life histories of particular galaxies, which argues
for as large a sample as possible.  Accordingly, we select two
galaxies from each environment to at least begin to explore such
variations, giving a final sample size of six galaxies.

Driven by these considerations, together with the observational
constraints of the William Herschel Telescope, the objects observed
for this project were NGC~3115 and NGC~7457 as isolated S0s, NGC~1023
and NGC~2768 in poor groups, and NGC~3384 and NGC~3489 which are both
in the richer Leo Group.  The basic properties of these galaxies are
summarised in Table~\ref{tab:gal1}, while the details of the PN.S
observations of the objects are provided in Table~\ref{tab:gal}.

\begin{table*}
\caption[Observational properties of the  galaxy sample (NED).]{Basic observational properties of the  galaxy sample. \label{tab:gal1}}
\centering  
\begin{tabular}{c c c c c c}
\hline\hline
 Name & Type &  D & cz & N{\sc{group}} & K \\
 $$ & $$ &  [Mpc] & [km/s] & $$ & [mag] \\
\hline
NGC~3115  &  S0-edge-on & $9.68$ & $663$ & $1$ & $ 5.88 \pm 0.017$ \\
NGC~7457  &  SA0-(rs) & $13.2$ & $812$ & $2$ & $8.192 \pm 0.024$ \\
NGC~2768  &  E6 & $22.4$ & $1373$ & $3$ & $6.997 \pm 0.031$ \\
NGC~1023  &  SB0-(rs) & $11.4$ & $637$ & $8$ & $6.238 \pm 0.021$ \\
NGC~3489  &  SAB0+(rs) & $12.1$ & $677$ & $14$ & $7.37 \pm 0.009$ \\
NGC~3384  &  SB0-(s) & $11.6$ & $704$ & $14$ & $6.75 \pm 0.019$ \\
\hline
\noindent
\end{tabular}
\begin{minipage}{18cm}
  Notes:
  Col. 1: Galaxy name.
  Col. 2: Morphological classification (NED).
  Col. 3: Distance to the galaxy \citep{Tonry}.
  Col. 4: Systemic velocity (NED).
  Col. 5: Number of members in the group the galaxy belongs to \citep{Fouque, DeVaucouleurs}.
  Col. 6: K-band apparent magnitudes (2MASS).  
\end{minipage}
\end{table*}

\begin{table*}
\caption[Observational properties of the  galaxy sample (PN.S).]{Basic observational properties of the  galaxy sample. \label{tab:gal}}  
\centering
\begin{tabular}{c c c c c  c c c c c c}
\hline\hline
Name & date & T{\sc{exp}}  &   seeing & filter & $\lambda$ & V{\sc{min}} & V{\sc{c}} & V{\sc{max}} & N{\sc{pn}}\\
  $$   &   $$   & [hrs]     &  [arcsec] & $$    & $[\mathring{A}]$ & [km/s] & [km/s] & [km/s] & $ $ \\
\hline
NGC~3115  &  $2007-2010$ & $10.6$    & $1.2$  & AB0 & $5026.00$ & $-168$ & $1150$ & $2467$ & $189$ \\
NGC~7457  &  $2001$       & $12.05$  & $1.15$ & B6  & $5026.74$ & $258$  & $1194$ & $2129$  & $113$ \\ 
NGC~2768  &  $2007$       & $15.0$   & $1.4$  & B0  & $5033.90$ & $686$  & $1623$ & $2560$  & $315$ \\ 
NGC~1023  &  $2007$       & $3/2.7$  & $1.0$  & AB0 & $5026.00$ & $-168$ & $1150$ & $2467$  & $204$ \\
NGC~3489  &  $2007-2011$ & $7.8$     & $1.4$  & AB0 & $5026.00$ & $-168$ & $1150$ & $2467$  & $60$ \\
NGC~3384  &  $2003-2011$ & $5.95$    & $1.6$  & AB0 & $5026.00$ & $-168$ & $1150$ & $2467$  & $93$ \\
\hline
\noalign{\smallskip}                                  
\end{tabular}
\begin{minipage}{18cm}
  Notes:
  Col. 1: Galaxy name.
  Col. 2: Years in which the  observations were carried out.
  Col. 3: Effective exposure time (exposure time normalised at $1"$ seeing). NGC~1023 has been observed in two fields of view and the tabled exposure times correspond to the East and West fields respectively. 
  Col. 4: Average seeing.
  Col. 5: Filter used.
  Col. 6: central wavelength of the filter.
  Col. 7/9: corresponding velocity window.
  Col. 8: central velocity.
  Col. 10: number of observed PNe.
\end{minipage}
\end{table*}

\subsection{Companion galaxies}
Since some possible channels for transformation into S0s involve
interactions with companion galaxies, we now examine the cases where
sample galaxies have relatively close neighbours.

NGC~1023 has a low surface-brightness companion, NGC~1023A, which lies
close to the main system \citep{Capaccioli}. In  their catalogue of PNe in NGC 2013, 
\citet{EDO} found that 20 of the 204 detected PNe were in fact associated with NGC 2013A.

\begin{figure*}[!t] 
\centering
\begin{tabular}{cc} 
\includegraphics[width=0.45\textwidth]{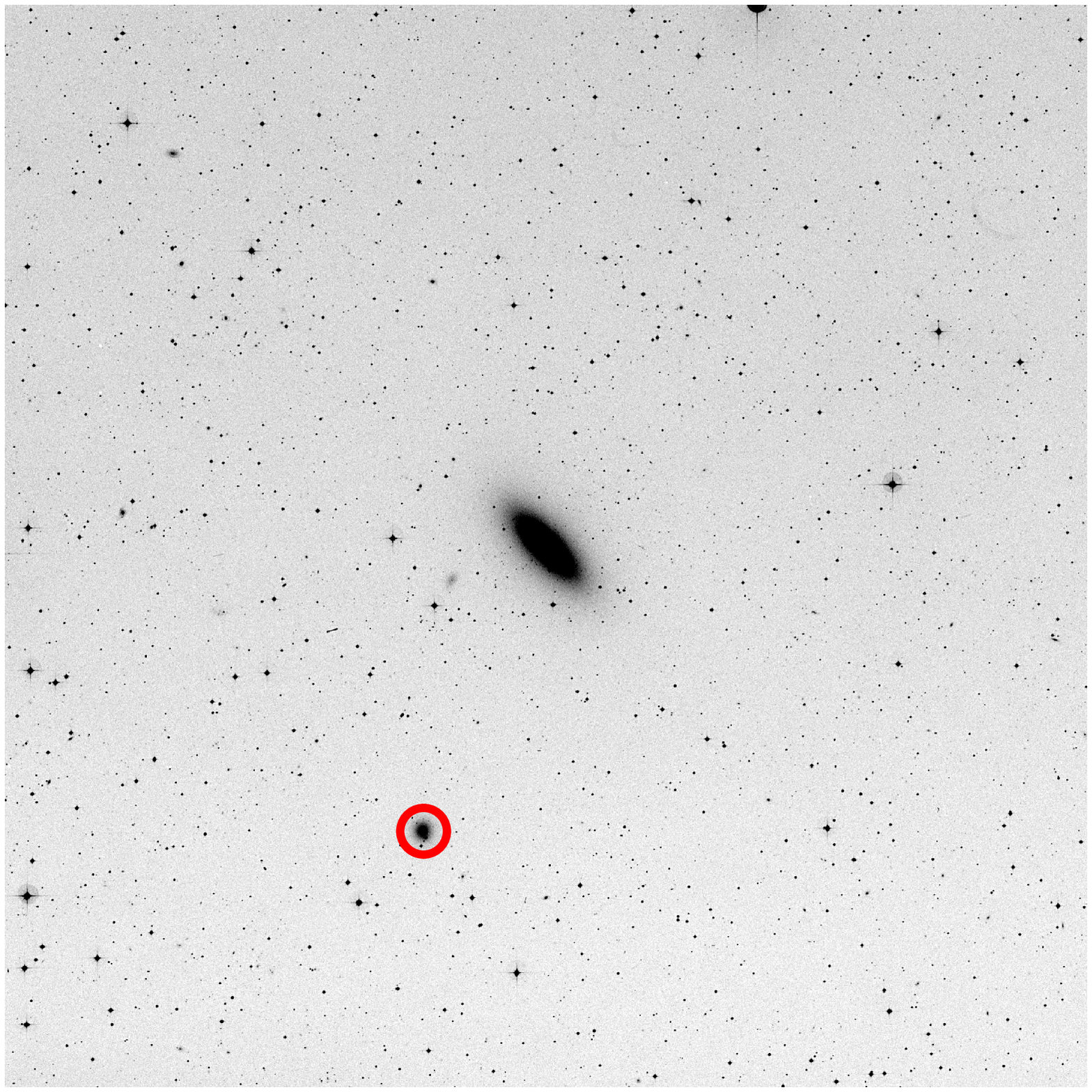}\includegraphics[width=0.45\textwidth]{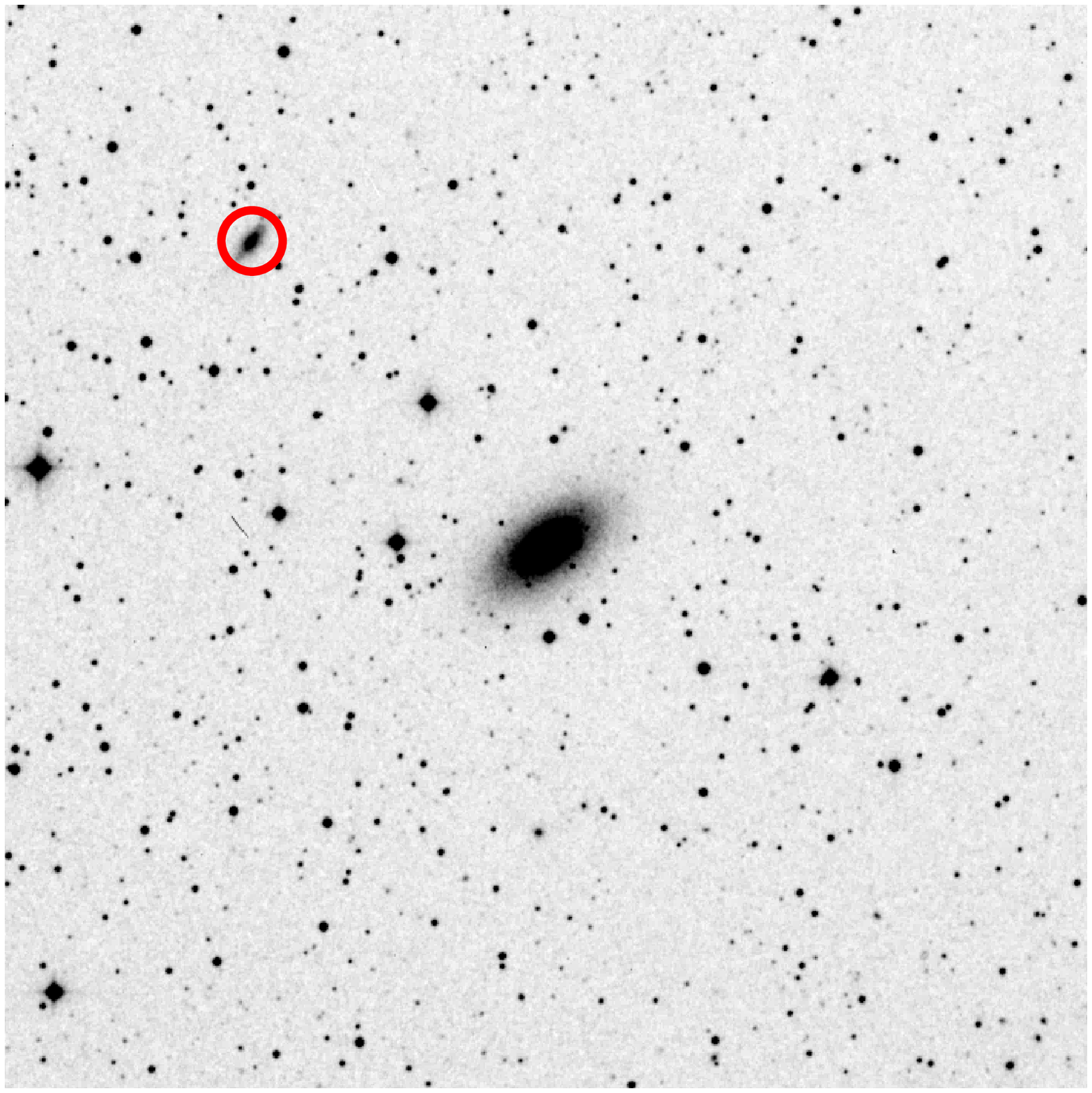}
\end{tabular}
\caption[Images of NGC~3115 and NGC~7457 and their companion galaxies.]{ESO Digital Sky Survey (DSS) images of NGC 3115 ({\it left panel}, $60'
\times 60'$) and NGC 7457 ({\it right panel} $14.8' \times 14.8'$) and their
companion galaxies, which are marked with red circles. North is up,
and East to the left in both panels. \label{fig:field}}
\end{figure*}

NGC~3115 is associated with MGC-1-26-12 (NGC~3115B) and MGC-1-26-21,
and it is surrounded by some H{\sc i} clouds. MGC-1-26-12 and
MGC-1-26-21 are two faint galaxies with $I$-band and $K$-band apparent
magnitudes of $14.26$ mag and $10.05$ mag respectively
\citep{Doyle}. MGC~1-26-21 is shown in the left panel of
Figure~\ref{fig:field}; its distance from the main galaxy is $17'$ and
it has a diameter of $70.3''$ \footnote{(2MASS,
  http://irsa.ipac.caltech.edu/)}.  MGC-1-26-12 is too far away from
the central galaxy to be seen in Figure~\ref{fig:field}, being at a
distance of $43'$. Its major axis measures $12.10''$
\citep{rc3}.

NGC~7457 is accompanied by UGC~12311 \citep{DeVaucouleurs}, which is
an irregular galaxy with a $K$-band apparent magnitude of $12.4$, and
a diameter of $23.6''$, and it lies at a distance of $5.7'$ from the
centre of NGC~7457.  It is shown in the right panel of
Figure~\ref{fig:field}.

With the possible exception of NGC~1023A, all of these companions are
faint and relatively distant, so are unlikely to have played a
significant role in the evolution of the sample galaxies.

\section{Data reduction and catalogues}
\label{sec:catalogues}

The data from PN.S are rather unusual in form, and require a
custom-built analysis package for their analysis.  The details of the
pipeline are presented in \citet{Nigel, Nigel2} and \citet{Merrett};
here we give a brief summary of the main steps, describe some
recent upgrades to its capabilities, and tabulate the resulting
catalogues of PNe.

\subsection{The Planetary Nebula Spectrograph and the reduction pipeline}
\label{subsec:pipeline}

The PN.S, described in detail in \citet{Nigel}, is a slitless
spectrograph that works using a counter-dispersed imagining technique
\citep{Fe1, Fe2, Fe3}. Two images of the same field are taken at the
same time through a slitless spectrograph equipped with a narrow-band
[OIII] filter to restrict the spectral range, and they are dispersed
in opposite directions. Thus, stars in the field appear in the image
as short segments of spectrum around 5007\,\AA\, while the PNe appear
as point sources due to their monochromatic emission. The exact
position of a PN  [\ion{O}{iii}] 5007\,\AA\ line in each image depends on
both its position on the sky and its observed wavelength, so matching
up the detected PNe in pairs from the images dispersed in opposite
directions makes it possible to obtain both their positions and
velocities in a single step.

To extract this information, we have developed a specialised reduction
pipeline for the instrument.  The first step in the reduction is to
create individual distortion-corrected images. This step includes the
creation of a bad pixel mask, cosmic ray removal, and finally the
undistortion of each image using arc lamp exposures obtained through a
calibrated grid mask.  As a second step, all images for a given galaxy
are aligned and stacked, weighted according to the seeing,
transparency etc.  Although stars appear elongated into trails in the
dispersed images, their centroids are still sufficiently well defined
for them to be used to perform an astrometric calibration, based on
the corresponding star coordinates in the USNO catalogue.  At this
point, the images are fully reduced, and we are ready to extract the
PNe data from them.

The procedure for extracting a catalogue of PNe from these images is
semi-automated, initially using Sextractor \citep{SEX} to find
potential objects. As explained above, PNe in the images appear as
point-like sources, since they emit only in the 5007\,\AA\ line within
the wavelength range selected by the filter.  With Sextractor,
experimentation shows that these real sources can be distinguished
from residual cosmic rays and other image defects by selecting only
those objects that comprise a group of at least 5 contiguous pixels
whose  individual fluxes are at least 0.7 times the rms of the background.

There is, however, still the possibility of contamination from
foreground stars and background galaxies.  Fortunately, such sources
are elongated along the dispersion direction due to their continuum
light emission, with the length of elongation dictated by the filter
bandpass.  To identify such objects automatically, we create two
further images.  The first is an unsharp mask formed by smoothing the
original spectral image with a $49 \times 49$ median filter, and
subtracting the smoothed image from the original.  The second is
obtained by convolving this median-subtracted image with an elliptical
Gaussian function elongated in the dispersion direction.  The
resulting images are almost the same for extended stellar trails, but
the point-like PNe are fainter in the smoothed version.  Thus, for
either arm, we can extract a catalogue that excludes any continuum
sources by matching up source detections, and selecting only sources
whose difference in magnitude between the median-subtracted image and
its smoothed counterpart is larger than 0.5 mag (again tuned by
experimentation).

\begin{figure*}[!t] 
\begin{tabular} {lr}
\includegraphics[width=1.\textwidth,height=1.\textwidth]{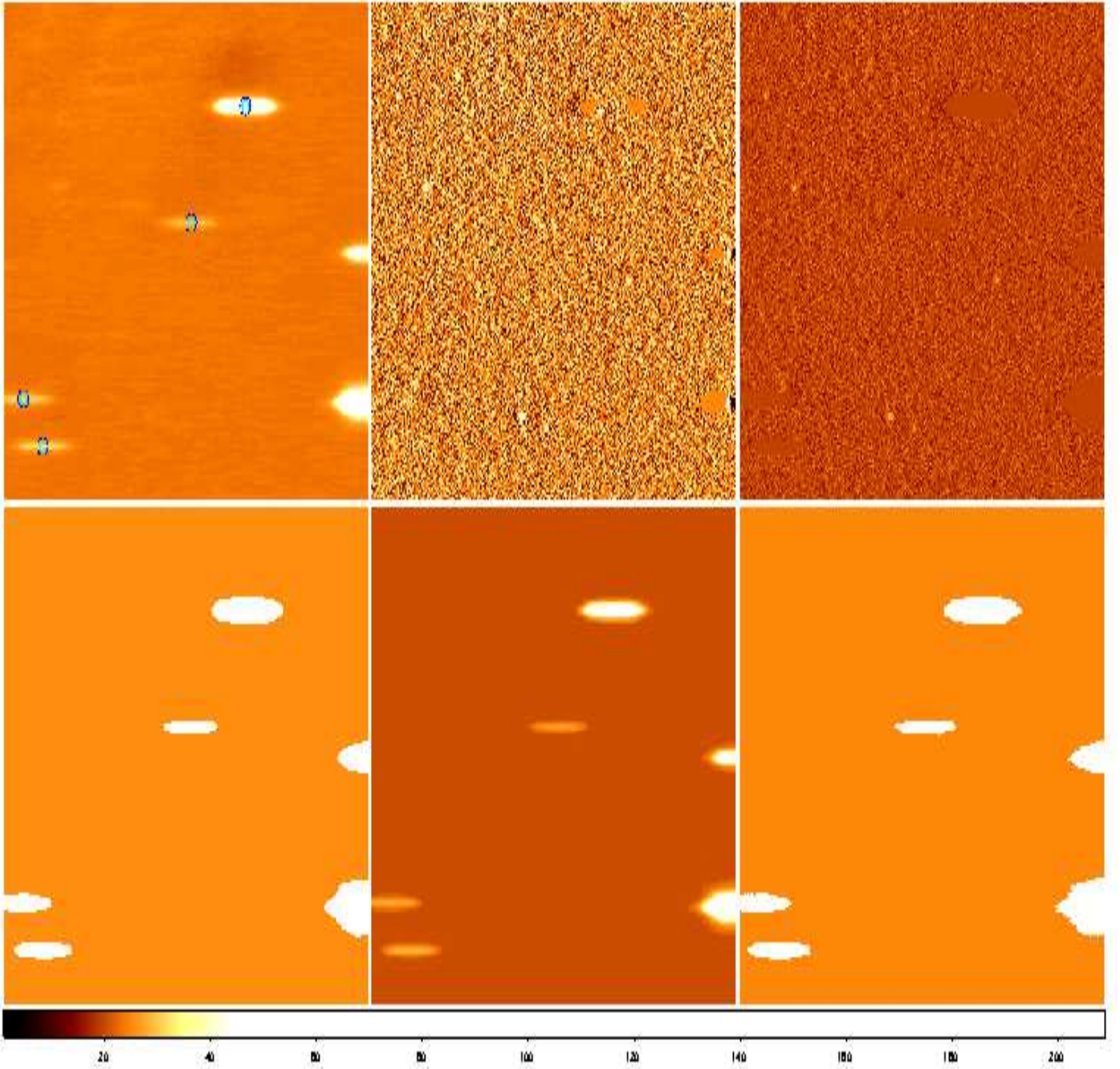}  
\end{tabular}
\caption[Different steps of the star trail removal routine.]{Different steps of the star trail removal routine, the top right panel show the resulting image. Refer to text for major details. \label{fig:removestar}}
\end{figure*}

Although this automated process, which we have used in our previous
pipeline processing, detects the vast majority of PNe, there are still
a number of missed detections when the object is projected near a star
trail.  We have therefore developed a further technique to subtract
out the star trails as an aid to detecting PNe, illustrated in
Figure~\ref{fig:removestar}.  We create a $25 \times 1$
median-filtered image (top left
panel) and subtract it from the science image, obtaining the median-subtracted image shown in the top middle
panel.  This image smooths away most of the star trails and leaves PNe
almost unaffected.  However, there are still residual features from
the brighter parts of the brightest star trails that could be misidentified as
PNe.  We must therefore create a mask to exclude these regions.  To do
so, we further apply a $81 \times 81$ median filter to the $25 \times
1$ median filtered image to obtain an image that picks out the stars
while removing the PNe.  To turn this image into a mask, we run
Sextractor on it and perform a morphological dilation on the
segmentation map shown in the bottom left panel, by blurring (bottom
middle panel) and thresholding (bottom right panel) the image.
Applying this mask, by replacing masked values with zero, then yields the final trail-free data frame shown
in the top right panel, which can then be pipeline processed in the
way described above, with no residual confusion due to foreground stars.

With this modification, the only PNe missed by the automated process
are those that lie very near the edge of the image, where the
effectiveness of median filtering is limited, or near the centre of
the target galaxy, where the bright continuum is noisy and difficult
to subtract fully.  However, the numbers are now sufficiently small
that any remaining PNe are easily added in manually.

Finally the catalogues obtained in this way for the two arms of the
spectrograph are matched up.  Any pairs that appear at the same
location in the undispersed direction to within two pixels, so have
consistent spatial coordinate, and lie within a few hundred pixels of
each other in the dispersed direction, so are consistent with the
likely offset due to their Doppler shifts in the counter-dispersed
frames, are identified as images of a single PN.  Any object that does
not appear in both arms is rejected as spurious.  As a final check,
each putative detection is assessed manually by three independent team
members, by checking its location in the original stacked images, and
an object is accepted in the catalogue if at least two classifiers
confirm that it is real.

The wavelength and astrometric solutions derived above then allow us
to obtain the mapping
\begin{equation}
 \{x_{l},y_{l},x_{r},y_{r}\} \rightarrowtail \{ x_{0},y_{0}, \lambda \} ,
\end{equation}
where $x_{l},y_{l},x_{r},y_{r}$ are the coordinates of the PNe
 in the left and right-dispersed images, $x_{0}$ and $y_{0}$
are the true sky coordinates and $\lambda$ is the  wavelength of emission
for each object.  The emission wavelength translates directly into a
line-of-sight velocity via the Doppler formula.

Errors are assigned to these velocities by propagating the errors in
the assigned values of $x_{l},y_{l},x_{r},y_{r}$ through the
transformation, further allowing for the uncertainty in the dispersion
solution itself.  A check on these uncertainties is provided by the
fact that the observation of NGC~3384 presented here overlaps with the
field containing NGC~3379 published by \citet{Nigel2}.  Matching the
catalogues results in 25 objects in both observations for which the
RMS difference between the data is $20\,{\rm km}\,{\rm s}^{-1}$,
implying an average error in each individual observation of $14\,{\rm
  km}\,{\rm s}^{-1}$, consistent with what the internal error analysis
assigned.

\subsection{The catalogues}
\label{subsec:givencatalogues}
The resulting catalogues of PNe are presented in
Tables~\ref{tab:3115cat}--\ref{tab:3384cat}.  
The data for NGC~1023
are not shown here, since they were published in \citet{EDO}. For completeness, Table~\ref{tab:3384} also
contains the 9 PNe identified as belonging to NGC~3384 in the adjacent
observation of NGC~3379 by \citet{Nigel2}.  
\begin{table}[!b]
\caption{Catalogue of PNe in NGC~3115. \label{tab:3115cat}}
\centering  
\begin{tabular}{ccccc}
\hline\hline
 ID & RA &  Dec  & V{\sc{helio}} & m  \\
 $$ & J2000 &  J2000 & [km/s] & [mag] \\
\hline
PNS-EPN-N3489-01 & 10:04:51 & -07:44:57 & 645.58 &   - \\
PNS-EPN-N3489-02 &  10:04:54.9 & -07:43:55.5 & 845.18 & 18.96 \\
PNS-EPN-N3489-03 & 10:04:55.5 & -07:43:37.4 & 307.08 & 19.1 \\
PNS-EPN-N3489-04 & 10:04:57.1 & -07:46:22.8 & 568.28 & 18.65 \\
. & . &. & .& . \\
. & . &. & .& . \\
. & . &. & .& . \\
PNS-EPN-N3489-188 & 10:05:36.2 & -07:41:28.1  & 804.58 & 17.56\\
PNS-EPN-N3489-189 & 10:05:36.5  & -07:41:49.7 & 722.08 & 19.17\\
\hline
\noalign{\smallskip}                                 
\end{tabular}
\begin{minipage}{8.4cm}
Notes: Details on the velocity and positional errors are given in Section\ref{subsec:pipeline}. Velocities, measured
from the  [\ion{O}{iii}] 5007\,\AA\ line, are corrected to heliocentric. The full catalogue is available at the CDS.
\end{minipage}
\end{table}
\begin{table}
\caption{Catalogue of PNe in NGC~7457. \label{tab:7457cat}}
\centering  
\begin{tabular}{c c c c c }
\hline\hline
 ID & RA &  Dec  & V{\sc{helio}} & m  \\
 $$ & J2000 &  J2000 & [km/s] & [mag] \\
\hline
PNS-EPN-N7457-001& 23:00:42.75& 30:10:10.7& 993.94 & 20.4 \\
PNS-EPN-N7457-002& 23:00:48.31& 30:08:58.2& 870.64& 20.24 \\
PNS-EPN-N7457-003& 23:00:51.14& 30:09:19& 886.94&  19.5 \\
PNS-EPN-N7457-004& 23:00:51.42& 30:09:44.1& 939.94& 19.69 \\
. & . &. & .& . \\
. & . &. & .& . \\
. & . &. & .& . \\
PNS-EPN-N7457-112& 23:01:8.34& 30:08:24.6& 755.84& 19.98 \\
PNS-EPN-N7457-113& 23:01:10.11& 30:06:58.2& 724.74& 19.83 \\
\hline
\noalign{\smallskip}                                  
\end{tabular}
\begin{minipage}{8.4cm}
Notes: Details on the velocity and positional errors are given   in Section\ref{subsec:pipeline}. Velocities, measured
from the  [\ion{O}{iii}] 5007\,\AA\ line, are corrected to heliocentric. The full catalogue is available at the CDS.
\end{minipage}
\end{table}
\begin{table}
\caption{Catalogue of PNe in NGC~2768. \label{tab:2768cat}}
\centering  
\begin{tabular}{c c c c c }
\hline\hline
 ID & RA &  Dec  & V{\sc{helio}} & m  \\
 $$ & J2000 &  J2000 & [km/s] & [mag] \\
\hline
PNS-EPN-N2768-01 & 09:10:56.13 & 60:03:21.9 & 1181.3 &  20.86 \\
PNS-EPN-N2768-02 &  09:10:59.59 &  60:02:13.7 &  1133.9 &  21.86 \\
PNS-EPN-N2768-03 &  09:10:59.86 &  60:03:38.5 &  1134.4 &  21.05 \\
PNS-EPN-N2768-04 &  09:11:1.34 &  60:03:3.1 &  1172.1 &  20.49 \\
. & . & . & . & . \\
. & . & . & . & . \\
. & . & . & . & . \\
PNS-EPN-N2768-314 &  09:12:15.08 & 60:03:8.9 & 1556.4 & 20.36 \\
PNS-EPN-N2768-315 &  09:12:15.22 & 60:00:3.6 & 1595.6 & 21.08 \\
\hline
\noalign{\smallskip}                                  
\end{tabular}
\begin{minipage}{8.4cm}
Notes: Details on the velocity and positional errors are given   in Section\ref{subsec:pipeline}. Velocities, measured
from the  [\ion{O}{iii}] 5007\,\AA\ line, are corrected to heliocentric. The full catalogue is available at the CDS.
\end{minipage}
\end{table}
As well as the positions
and line-of-sight velocities of the PNe, these tables also show the
instrumental magnitude of each object, as derived from the images;
since we only use these magnitudes to assess the strength of the
detection and to investigate the shape of the planetary nebula
luminosity function (see Section~\ref{sec:PNe surface density}), no
photometric calibration is required.
\begin{table}
\caption{Catalogue of PNe in NGC~3489. \label{tab:3489cat}}
\centering  
\begin{tabular}{c c c c c }
\hline\hline
 ID & RA &  Dec  & V{\sc{helio}} & m \\
 $$ & J2000 &  J2000 & [km/s] & [mag] \\
\hline
PNS-EPN-N3489-01 & 11:00:09.1 & 13:53:21.3 & 660.52 & 16.49 \\
PNS-EPN-N3489-02 & 11:00:09.2 & 13:55:26.3 & 663.62 & 16.59 \\
PNS-EPN-N3489-03 & 11:00:09.6  & 13:51:25.7 & 1012.62 & 19.17 \\
PNS-EPN-N3489-04 & 11:00:10.4 & 13:53:45.1 & 1418.92 & 16.49 \\
. & . &. & .& . \\
. & . &. & .& . \\
. & . &. & .& . \\
PNS-EPN-N3489-59 & 11:00:23.7 &  13:54:43.8 &  742.12 & 15.82 \\
PNS-EPN-N3489-60 & 11:00:27.4 &  13:54:41.6 &  591.12 &  16.62 \\
\hline
\noindent{\smallskip}
\end{tabular}
\begin{minipage}{8.4cm}
Notes: Details on the velocity and positional errors are given   in Section\ref{subsec:pipeline}. Velocities, measured
from the  [\ion{O}{iii}] 5007\,\AA\ line, are corrected to heliocentric. The full catalogue is available at the CDS.
\end{minipage}
\end{table}
\begin{table}
\caption{Catalogue of PNe in NGC~3384. \label{tab:3384cat}}
\centering  
\begin{tabular}{c c c c c }
\hline\hline
 ID & RA &  Dec  & V{\sc{helio}} & m  \\
 $$ & J2000 &  J2000 & [km/s] & [mag] \\
\hline
PNS-EPN-N3384-01& 10:47:54.8& 12:36:48.5& 822.6& 18.75\\
PNS-EPN-N3384-02& 10:47:54.9& 12:36:15.8& 1038.7& 19.01 \\
PNS-EPN-N3384-03& 10:47:55.1& 12:36:16.8& 868.1& 18.98 \\
PNS-EPN-N3384-04& 10:47:57.1& 12:36:02.8& 908.7& 18.86 \\
. & . &. & .& . \\
. & . &. & .& . \\
. & . &. & .& . \\
PNS-EPN-N3384-94& 10:48:31.2& 12:39:02.6& 646.9& 18.41 \\
PNS-EPN-N3384-95& 10:48:33.2& 12:36:26.9 &  629& 18.23 \\
\hline
\noindent{\smallskip}
\end{tabular}
\begin{minipage}{8.4cm}
Notes: Details on the velocity and positional errors are given   in Section\ref{subsec:pipeline}. Velocities, measured
from the  [\ion{O}{iii}] 5007\,\AA\ line, are corrected to heliocentric. The full catalogue is available at the CDS.
\end{minipage}
\label{tab:3384}
\end{table}

\section{The PNe Spatial Distribution and its Normalisation}
\label{sec:PNe surface density}
As discussed in the Introduction, one concern with using PNe as
tracers is that they may not be representative of the whole stellar
population, but rather a subcomponent, which may introduce a bias in
the dynamical analysis.  This issue has already been thoroughly
addressed in the context of ellipticals \citep{Coccato,Napolitano2,Napolitano3}, but now we need to establish
the same point for S0s.  Indeed, similar findings for S0s and
ellipticals would not only confirm the effectiveness of the technique,
but might also allow us to draw connections between the two types of
galaxy.

\subsection{Spatial distribution}
The simplest test we can make is to compare the spatial distribution
of PNe to that of the underlying stellar population: if PNe are 
effectively a random sampling of stars, then, to within a
normalisation factor, the two should be the same.

\begin{figure*}[!t] 
\begin{tabular} {lr}
\includegraphics[width=0.45\textwidth]{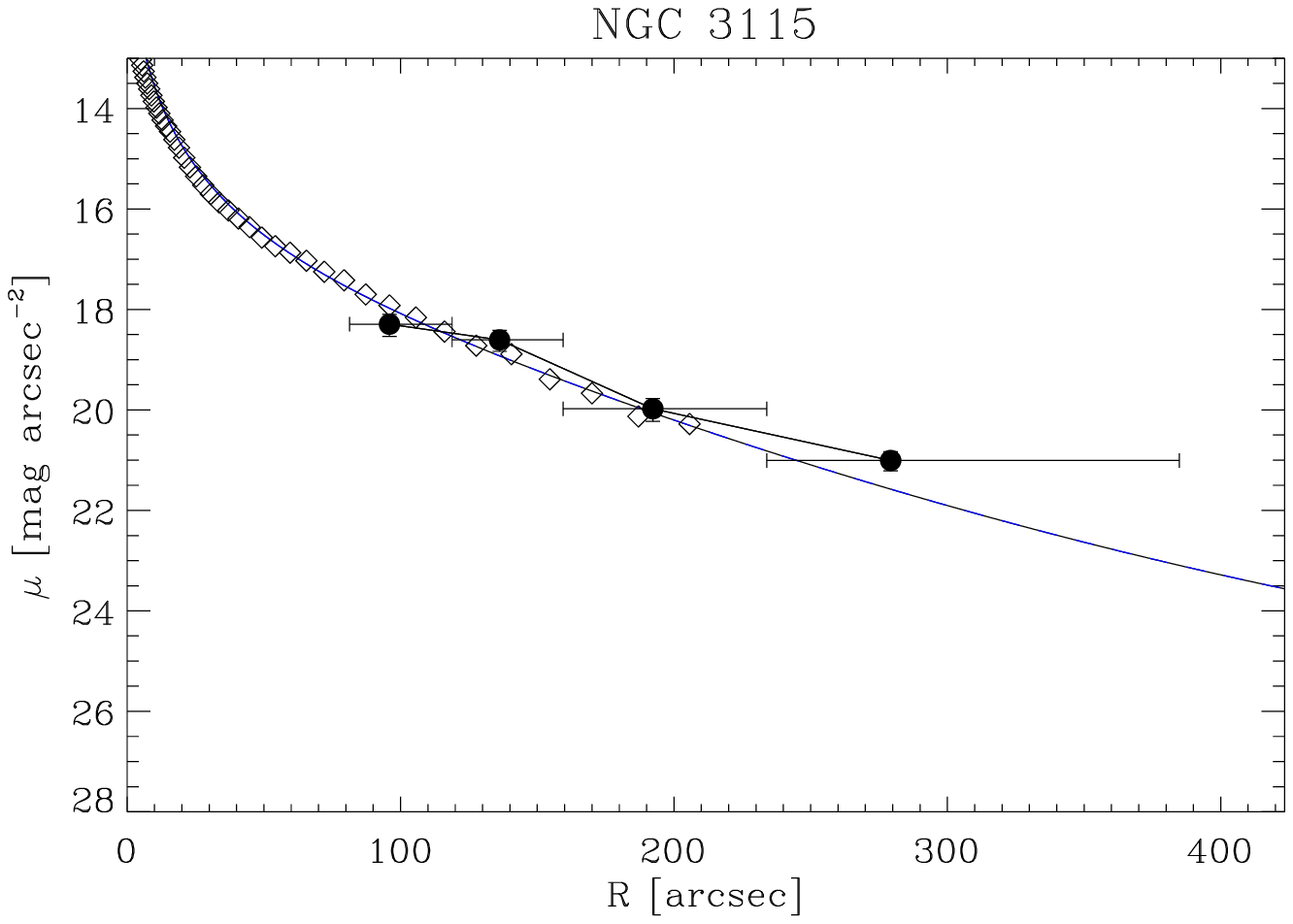} &  \includegraphics[width=0.45\textwidth]{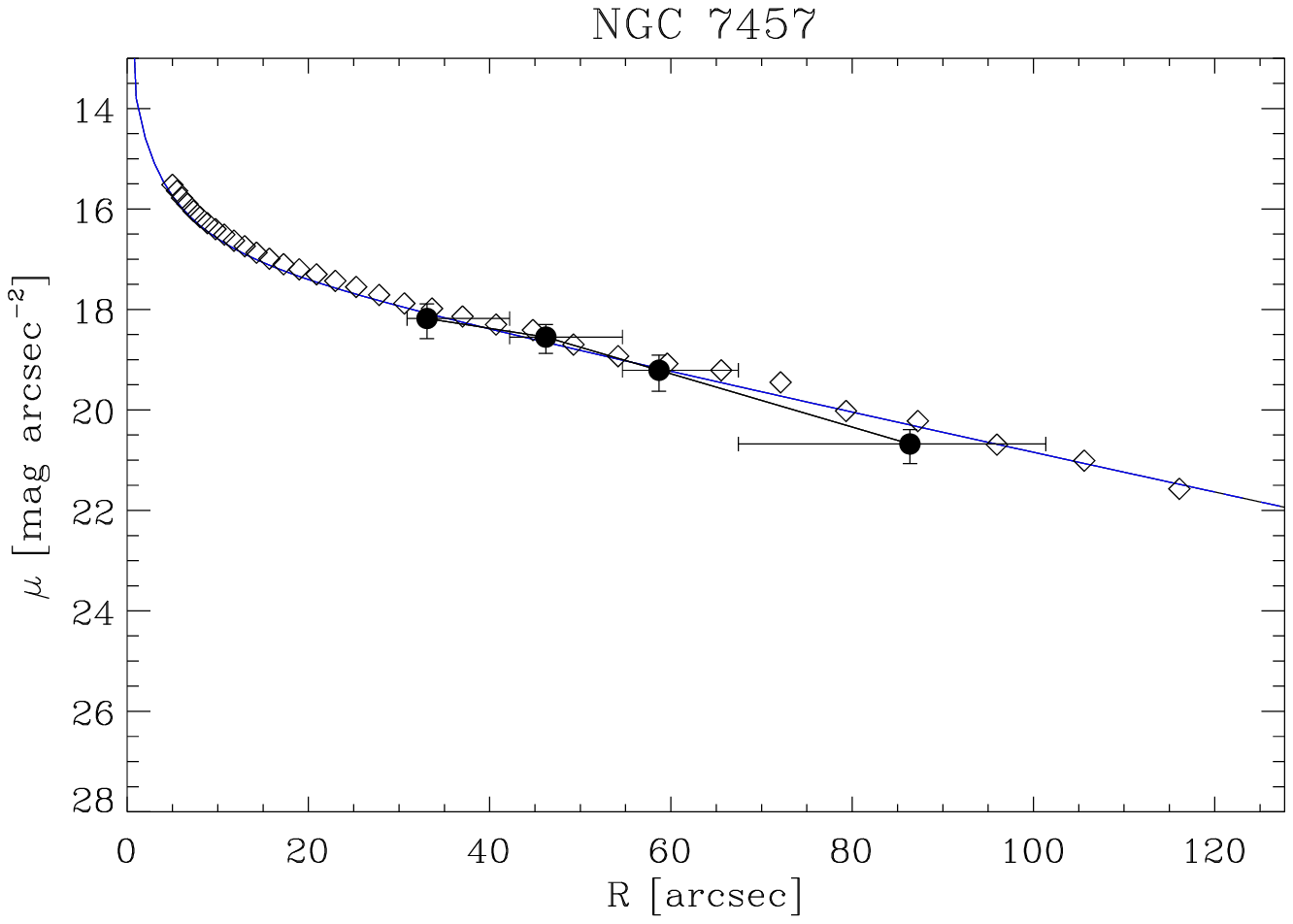}\\
\includegraphics[width=0.45\textwidth]{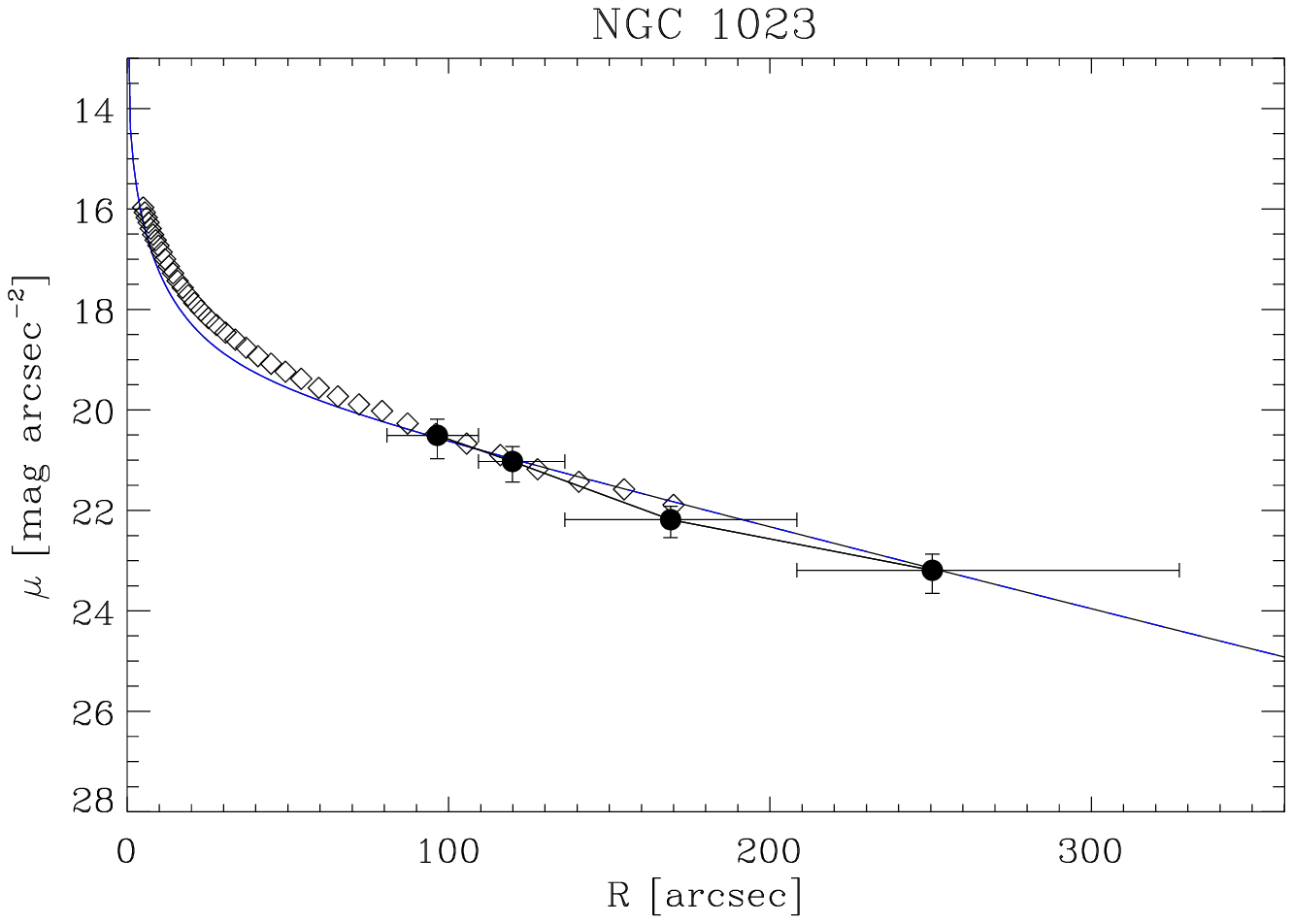} &  \includegraphics[width=0.45\textwidth]{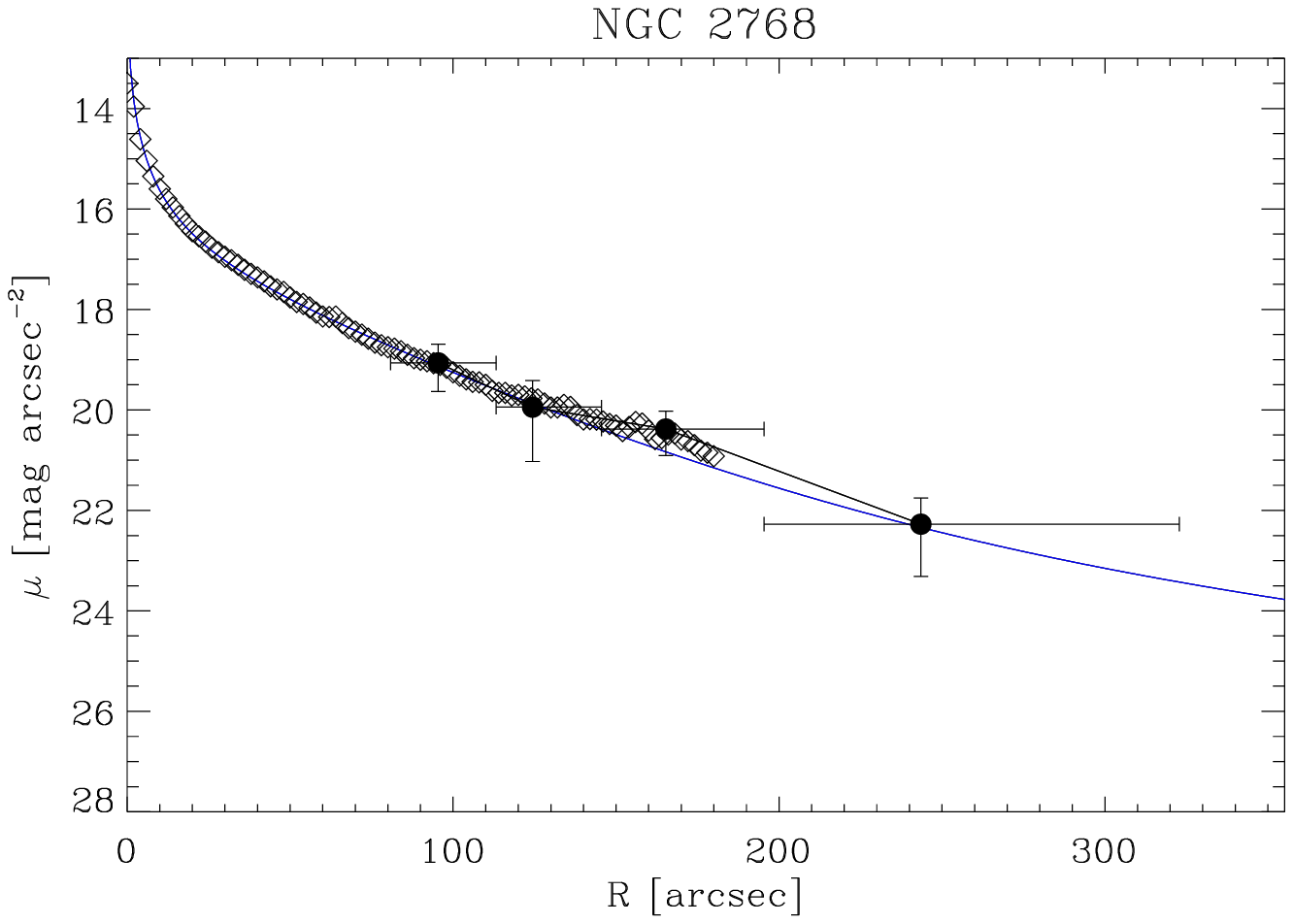}\\
\includegraphics[width=0.45\textwidth]{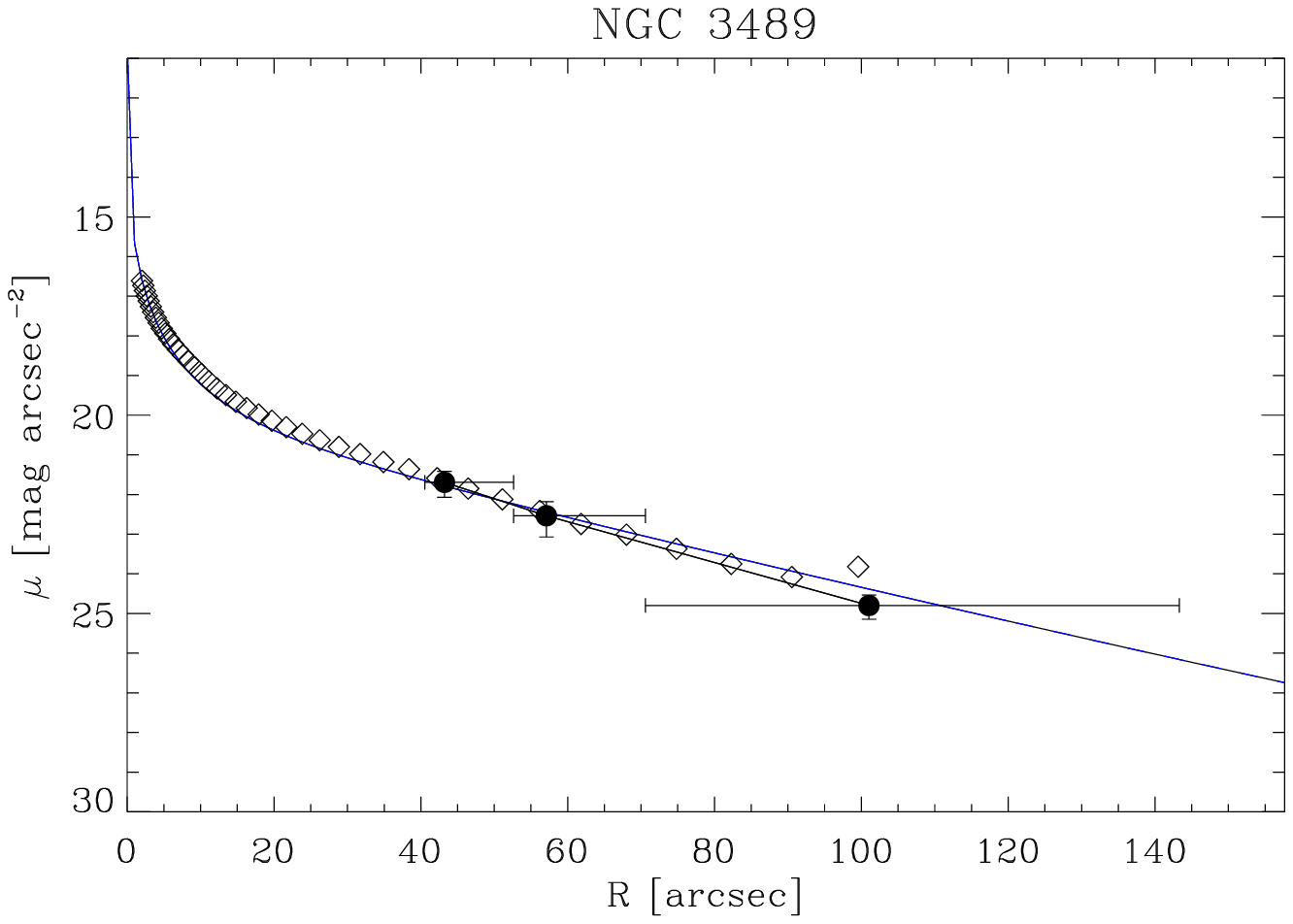} &  \includegraphics[width=0.45\textwidth]{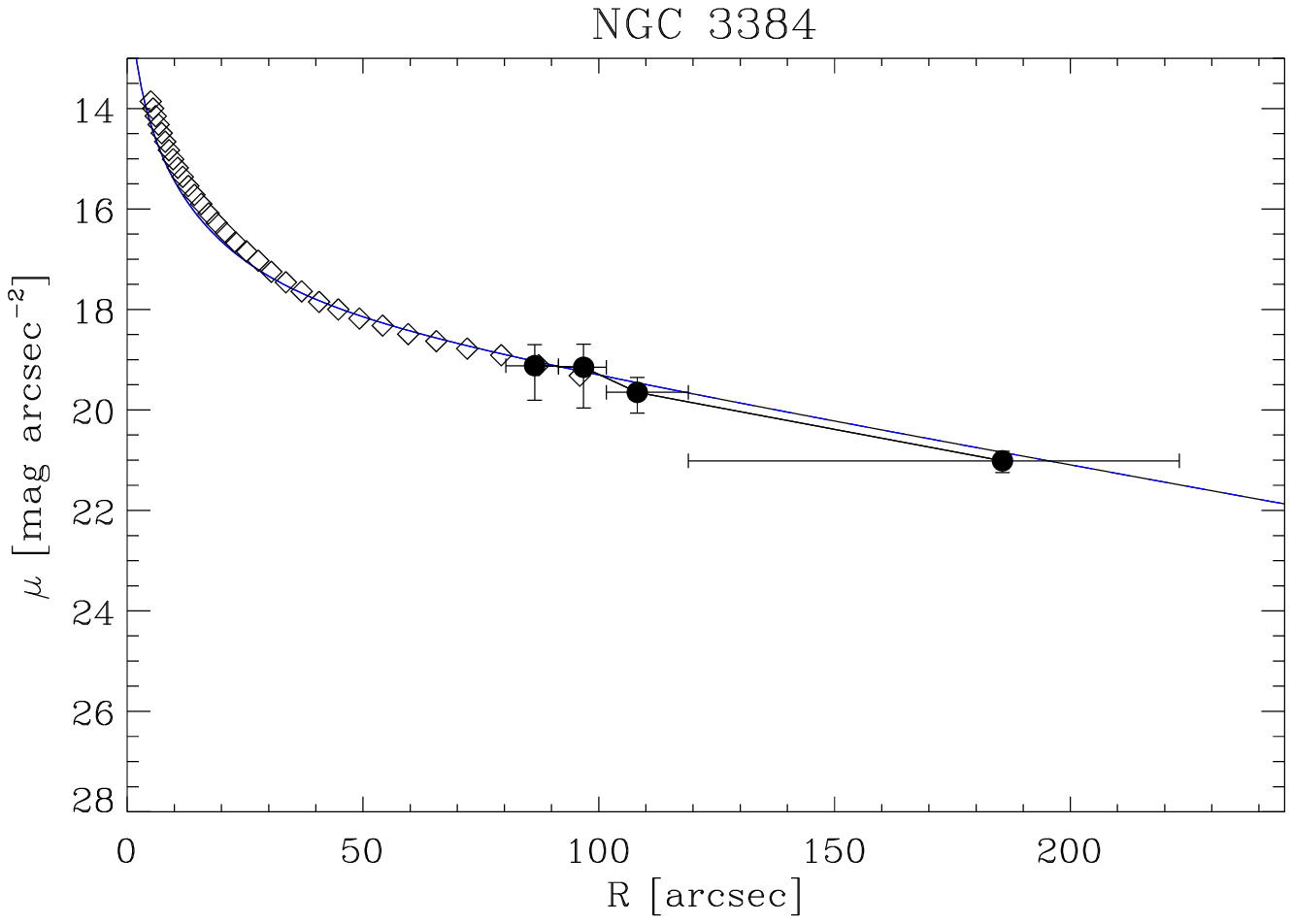}\\
\end{tabular}
\caption[PNe number counts in comparison with surface brightness profile.]{ Surface density profiles of PNe and stars. Open diamonds: surface
brightness profiles obtained with ellipse fitting of the galaxy image. Filled circles: PNe number
density, computed as $-2.5 Log Nc/A$, where $Nc$ is the completeness
corrected number of PNe with $m<m_{80}$ in the radial bin, and $A$ is
the area of the circular anulus.  Blue lines represent the
extrapolation of the sersic + exponential disk fit to the stellar
surface brighness data.   \label{fig:pneandstarlight}}
\end{figure*}

We have therefore fitted elliptical isophotes to images of the
galaxies in the sample, and plotted the surface brightness profiles
extracted from these fits in Figure~\ref{fig:pneandstarlight}.  In
most instances, the images used were the $K$-band observations from the
2MASS survey \citep{Skrutskie+06}.  In the case of NGC~3489, the galaxy
is too faint to be reliably fitted using this data, but fortunately
this galaxy lies in the region surveyed by SDSS \citep{York+00}, so
we used their $z$-band image as a substitute: since we are only
interested in matching the shape of the brightness profile, this red
band traces the old stellar population as well as the $K$-band, and no
colour correction is required.  Finally, in the case of NGC~1023 we
made use of a deep $R$-band image that we had previously analysed to
determine the size of the region compromised by light from NGC~1023A,
to exclude it from the analysis \citep{Ari}.

Typically, the PNe extend to larger radii than the reliable
photometric fits shown in Figure~\ref{fig:pneandstarlight}.  In order
to extrapolate the comparison to these large radii, we have also
fitted the images with a simple two-component disk+bulge model using
GALFIT \citep{Peng}, and obtained the total surface brightness profile, resulting in the lines shown in
Figure~\ref{fig:pneandstarlight}.  

\begin{figure}[!t] 
\begin{tabular} {lcr}
\includegraphics[width=0.45\textwidth]{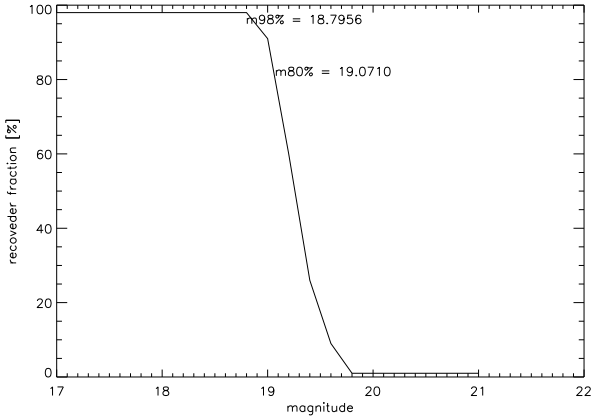} 
\end{tabular}
\caption[Main steps of the incompleteness correction procedure for NGC~3115]{Recovered number of point like sources as a function of  magnitude. The simulated PNe are added on an artificial image that mimic the sky and CCD noise, the recovered number decreases for fainter objects, see text for further details.   \label{fig:compl1}}
\end{figure}

\begin{figure}[!t]
\begin{tabular} {lcr}
\includegraphics[width=0.45\textwidth]{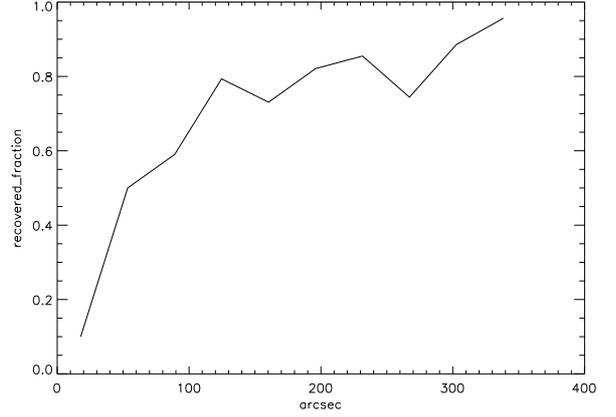} 
\end{tabular}
\caption[Main steps of the incompleteness correction procedure for NGC~3115]{Number of detected point like sources with $m>m_{80}$ as a function of radius; the simulated objects are added to the science image. The incompleteness at small radii is caused by the bright continuum of the galaxy, see text for further details.   \label{fig:compl2}}
\end{figure}

In order to compare these profiles to the PNe data, we need to correct
the latter for incompleteness.  Since PNe are primarily lost against
the bright central parts of the galaxy, failing to correct for this
effect would grossly distort their spatial profile.  We carry out this
correction in two stages, following a process developed by
\citet{Coccato}.  First, we randomly place simulated PNe of varying
magnitudes in model images that reproduce the noise properties of the
real data, to establish the brightness level to which PNe can in
general be reliably detected.  We quantify this limits by $m_{80}$,
the magnitude limit at which 80\% of the simulated PNe are recovered.
An illustration of this process is shown in 
Figure~\ref{fig:compl1}.  Next, we populate the real science data with
simulated PNe drawn from the known planetary nebula luminosity
function (PNLF), brighter than an $m_{80}$ completeness limit, and see
what fraction are lost against the bright galaxy continuum, star
trails, etc, as a function of position, as illustrated in Figure~\ref{fig:compl2}.  Finally, we correct for this
incompleteness in the number counts of PNe as a function of radius,
plotted in Figure~\ref{fig:pneandstarlight}.

In the central parts of these galaxies, incompleteness is too great to
be corrected reliably, but, as Figure~\ref{fig:pneandstarlight} shows,
in the region of overlap with the photometric data, out into the
extrapolation of the model fit, the agreement between the profiles of
PNe and the stellar continuum is very good, typically over at least a
factor of ten in brightness.  There is therefore no evidence that the
PNe are anything other than a fair sample of the stellar population.

\begin{figure}[!t]
\begin{tabular} {lcr}
\includegraphics[width=0.45\textwidth]{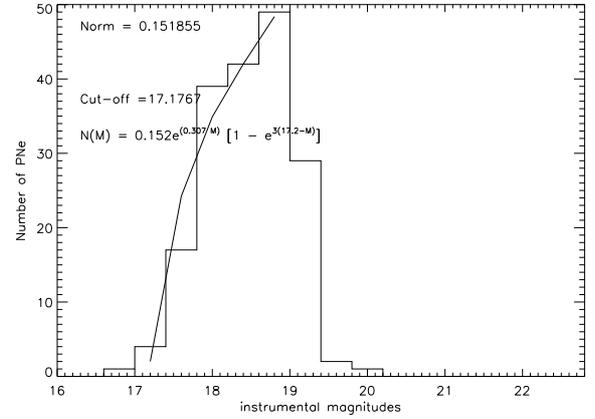}
\end{tabular}
\caption[Main steps of the incompleteness correction procedure for NGC~3115]{The observed PNLF and the fitted analytic function \citep{Ciardullo1} of NGC~3115, see text for further details.   \label{fig:compl3}}
\end{figure}

\subsection{The specific frequency of PNe}
\label{sec:specific frequency}
Having established that the proportionality between PNe numbers and
stellar luminosity seems to be valid, we can now calculate the
proportionality constant.  This factor, usually designated $\alpha$
\citep{Jacoby80}, is of interest because it relates to the
stellar evolution of the particular population.  So, for example,
assessing how similar it is between ellipticals and S0s provides an
independent route for determining how closely related these objects
are, in this case in terms of the evolution of their stellar
populations.  Such a similarity would also add further justification
for the view that PNe in S0s have a similar relation to the underlying
stellar population as those in ellipticals, and hence that they can be
used as similar kinematic tracers.

In fact, there is already evidence that this proportionality varies
in interesting ways with the global properties of the galaxy.  Although
$\alpha$ is close to universal in value, it does vary systematically
with colour \citep{Hui93}, with the reddest ellipticals being poorer in PNe per unit galaxy luminosity than spirals \citep{Buzzoni}.  This is somewhat
counter-intuitive, as one might expect a star-forming spiral to have
more of its luminosity provided by massive young stars that will never
form PNe \citep{Buzzoni}.  However, there are other ways in which the
$\alpha$ parameter might be reduced.  For example, if PNe have
systematically shorter lifetimes in elliptical galaxies, one would
expect to see fewer of them in any snapshot view.  Alternatively, some
stars with particular chemical properties might avoid becoming PNe at
all, perhaps by losing their envelopes at an earlier stage.  Support
for this suggestion comes from the rather tight anti-correlation
observed between $\alpha$ and the amount of far ultra-violet (FUV)
emission from the galaxy: in an old stellar population, FUV emission could
well arise from AGB stars that have already lost their envelopes,
thus avoiding later evolution into PNe \citep{Buzzoni}.

We therefore now calculate the $\alpha$ parameter for the sample S0
galaxies, to see how they compare to those already determined for
ellipticals \citep{Coccato}.  The value of $\alpha$ obviously depends
on how far down the PNLF one counts PNe away from the brightest
magnitude observed.  This brightest magnitude is well-defined by a
sharp cut-off in the PNLF at a magnitude $m^*$, which we can measure
by fitting the \citet{Ciardullo1} analytic approximation to the data,
as illustrated in  Figure~\ref{fig:compl3} (kept in
instrumental magnitudes since an absolute value is not required).  The
value of $\alpha$ then depends on how far below $m^*$ one counts PNe.
For a direct comparison with the ellipticals, we follow the same
definition as \citet{Coccato}, which involves counting the number of
PNe with apparent magnitudes down to $m^* + 1$.  We then calculate
$\alpha_{B,1.0}$, which is the ratio of this number of PNe to the
total $B$-band luminosity over the range of radii that PNe have been
observed.  The radial range used for each galaxy in this survey is
listed in Table~\ref{tab:alpha}.

In practice, our data generally probe to magnitudes
significantly fainter than $m^* + 1$: Table \ref{tab:alpha} lists the values
of $m^{*}-m_{80}$ which quantifies how many magnitudes below $m^*$ the
data remain reasonably complete.  In order to use all this information in
determining $\alpha$, we calculate the constant of proportionality
counting PNe all the way down to this limiting
magnitude, $\alpha_{B,m^{*}-m_{80}}$, and then use the known shape of
the PNLF to correct its value to the adopted common metric quantity, 
\begin{equation}
 \alpha_{B,1.0}= \alpha_{B,m^{*}-m_{80}} \frac{\int_{m^{*}}^{m^{*+1}} F(m^{*},m^{'})dm^{'}}
                                         {\int_{m^{*}}^{m_{80}} F(m^{*},m^{'})dm^{'}},
\end{equation} 
where $F(m^{*},m)$ is the universal analytic approximation to the
PNLF found by \citet{Ciardullo1}. The values for $\alpha_{B,1.0}$
determined in this way are listed in Table~\ref{tab:alpha}.

\begin{table*}
\caption[$\alpha_{B,1.0}$ parameter and UV excess for the  galaxy sample.]{$\alpha_{B,1.0}$ parameter and UV excess for the  galaxy sample.\label{tab:alpha}}  
\centering
\begin{tabular}{ccccccccc}
\hline\hline
Name  & N{\sc{pn}} & N{\sc{pn,corr}} & R{\sc{min}} & R{\sc{max}}  & $m_{*}-m_{80}$ & $\alpha_{B,1.0}$ & (FUV-V) & B{\sc{t}}\\ 
 $ $  & $ $   &       $ $    & [arcs] & [arcs] & [mag]   & $10^{9} \cdot L_{\odot}$ & [mag] & [mag] \\
\hline
\hline
\noindent{\smallskip}
NGC~3115  &  $182$ & $186$  & $45.9$  & $389.8$   & $-3.0$ & $1.0 \pm 0.2$ & $7.6 \pm 0.1$  & $-20.03 \pm 0.13$ \\ 
NGC~7457  &  $94$  & $156$  & $10.6$  & $115.4$   & $-1.9$ & $8.4 \pm 3.0$ & $8.1 \pm 0.1$  & $-18.69 \pm 0.32$ \\
NGC~2768  &  $201$ & $289$  & $25.1$  & $328.0$   & $-1.6$ & $2.1 \pm 0.6$ & $7.6 \pm 0.1$  & $-20.89 \pm 0.34$ \\
NGC~1023  &  $133$ & $236$  & $21.7$  & $332.4$   & $-2.5$ & $2.5 \pm 0.6$ & $7.6 \pm 0.1$  & $-19.10 \pm 0.28$ \\
NGC~3489  &  $50$  & $75$   & $20.3$ & $146.0$   & $-3.3$ & $1.9 \pm 0.8$ & $6.6 \pm 0.1$  & $-20.05 \pm 0.22$ \\
NGC~3384  &  $75$  & $77$   & $43.6$  & $228.1$   & $-2.6$ & $2.3 \pm 0.9$ & $7.5 \pm 0.1$  & $-19.41 \pm 0.19$ \\
\hline
\noindent{\smallskip}
\end{tabular}
\begin{minipage}{18cm}
  Notes:
  Col. 1: Galaxy name.
  Col. 2: Number of PNe with $m < m_{80}$. 
  Col. 3: Number of PNe with $m < m_{80}$  once corrected for completeness. 
  Col. 4: Galactocentric radius of the innermost detected PN.  
  Col. 5: Galactocentric radius of the outermost detected PN.
  Col. 6: Difference between the PNLF bright cut-off ($m_*$) and $m_{80}$ 
          (see Section \ref{sec:specific frequency}).  
  Col. 7:  $\alpha$ parameter in the $B$-band, integrating the PNLF from $m_*$ to $m_{*} + 1$ mag
          (see Section \ref{sec:specific frequency}).
  Col. 8: UV excess, from the Galaxy Evolution Explorer (GALEX).
  Col. 9: $B$-band absolute magnitude, from the apparent magnitudes
in the RC3 catalogue and the distance moduli of \citet{Tonry}, shifted
by $-0.16$ magnitudes, see text for details.
\end{minipage}
\end{table*}

\begin{figure*}[!t] 
\begin{tabular} {c}
\includegraphics[width=0.9\textwidth]{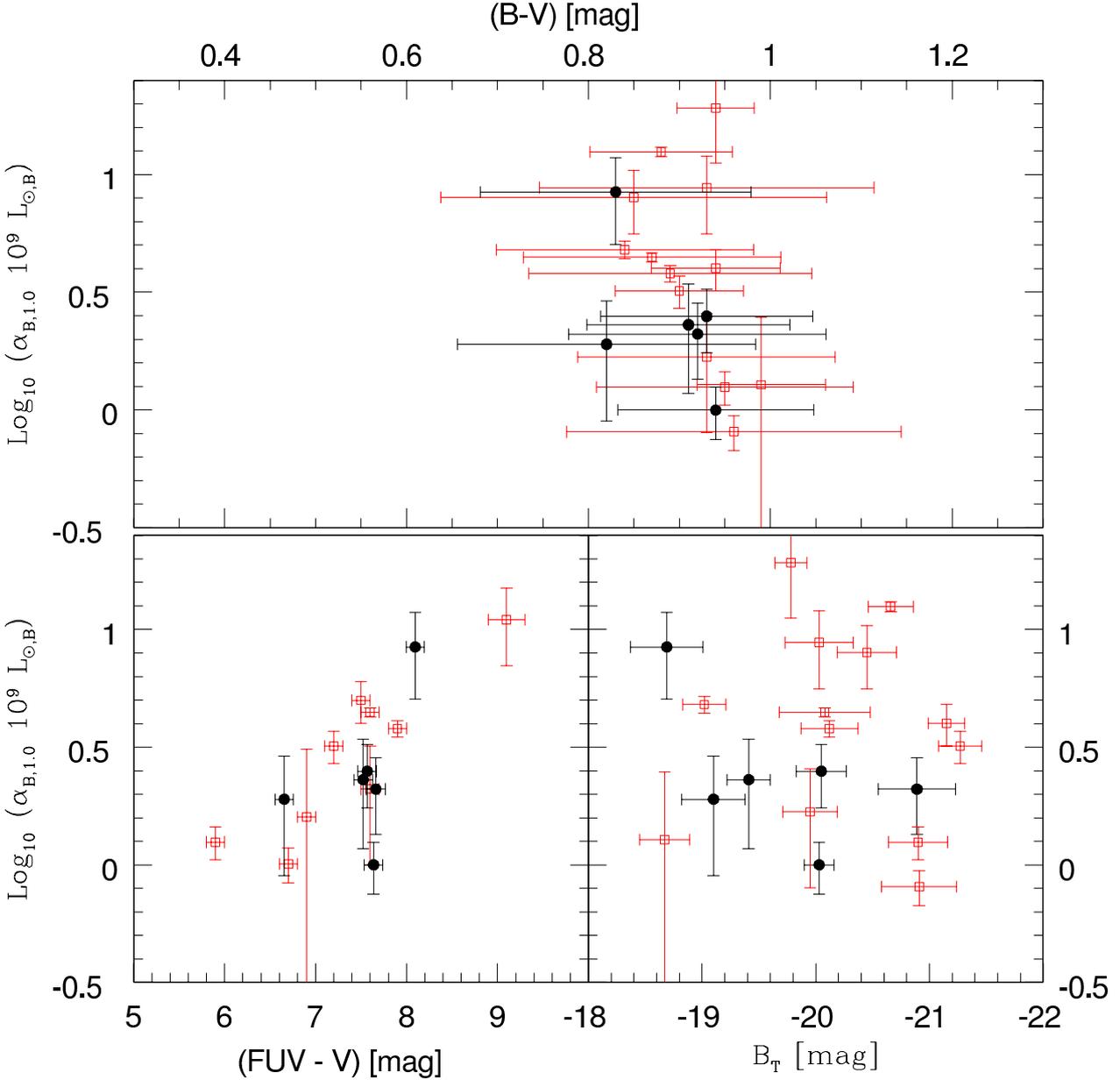} 
\end{tabular}
\caption[Alpha parameter.]{Relationships for elliptical and S0 galaxies between   $\alpha_{B,1.0}$ parameter and  UV excess (bottom left panel), absolute $B$-band magnitude (bottom-right panel) and (B-V) colour (upper panel).  Red open squares show the values of the elliptical sample presented in \citet{Coccato}, while filled black circles correspond to  S0s from the present study. \label{fig:alphavsE}}
\end{figure*}

Having calculated these $\alpha$ values, we can now compare them to
the values determined in the same way for ellipticals by
\citet{Coccato}.  Figure~\ref{fig:alphavsE} shows the resulting
comparisons when combining $\alpha$ values with optical colours, FUV
excess and absolute magnitude.  Plotted against optical colour, the
S0s $\alpha$ values are indistinguishable from ellipticals, and seem
to show the same trend described above that bluer galaxies have higher
specific frequencies of PNe.  Similarly, when we investigate their FUV
fluxes, obtained from the Galaxy Evolution Explorer
(GALEX)\footnote{GALEX data are available at:
  http://galex.stsci.edu/GR6/} and corrected for extinction using the
relation $A_{FUV}=8.376E(B-V)$ \citep{Wyder2005}, we find that the S0s
lie in the same part of the plot as ellipticals, although the range of
FUV excesses for the S0s is too small to really see if they follow the
same strong trend.  Finally, we compare $\alpha$ to the $B$-band
absolute magnitudes, which were obtained from the apparent magnitudes
in the RC3 catalogue and the distance moduli of \citet{Tonry}, shifted
by $-0.16$ magnitudes to take into account the newer Cepheid
zero-point of \citet{Freedman}, for consistency with \citet{Coccato}.
There is an interesting suggestion of a correlation between $\alpha$ and $FUV-V$ for S0s in Figure~\ref{fig:alphavsE}, 
which is strengthened by the fact that two of the elliptical
galaxies that lie closest to the S0s (NGC~4564 and NGC~3377) were found by  \citet{Coccato}
 to have disk like kinematics, and thus may well be S0s
themselves.  However, the numbers are still
sufficiently small that the only firm conclusion is that S0s lie in the
same region of these plots as ellipticals, and hence that the stellar
evolutionary properties that they probe are indistinguishable, adding
further confidence to the use of PNe as a common stellar kinematic
probe in these galaxy types.

\section{Velocity maps and kinematics}
\label{sec:velocity profile}

\begin{figure*}[p] 
\centering
\begin{tabular} {lr}
\vspace{10pt}
\includegraphics[width=0.4\textwidth,height=0.4\textwidth]{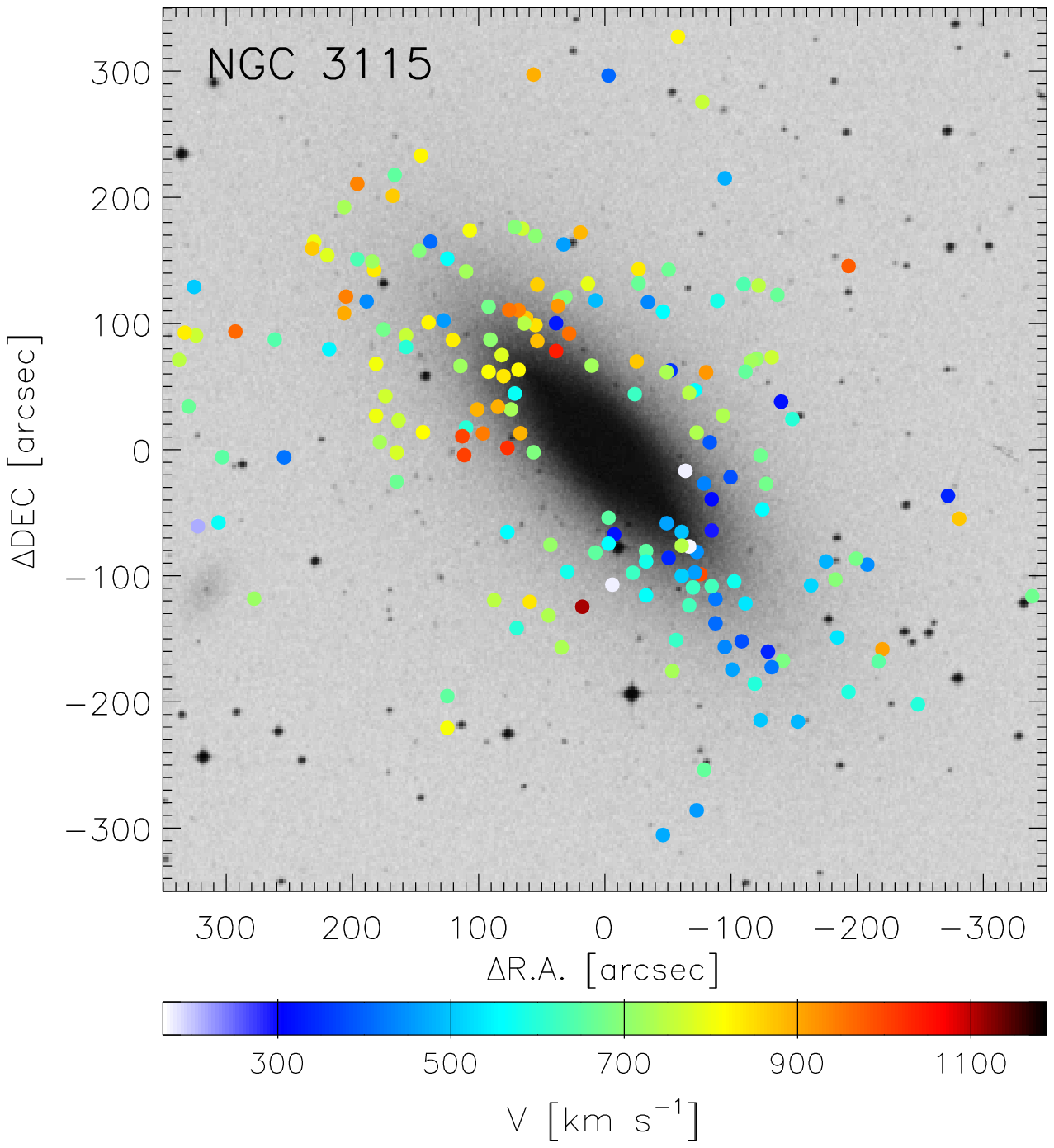} &  \includegraphics[width=0.4\textwidth,height=0.4\textwidth]{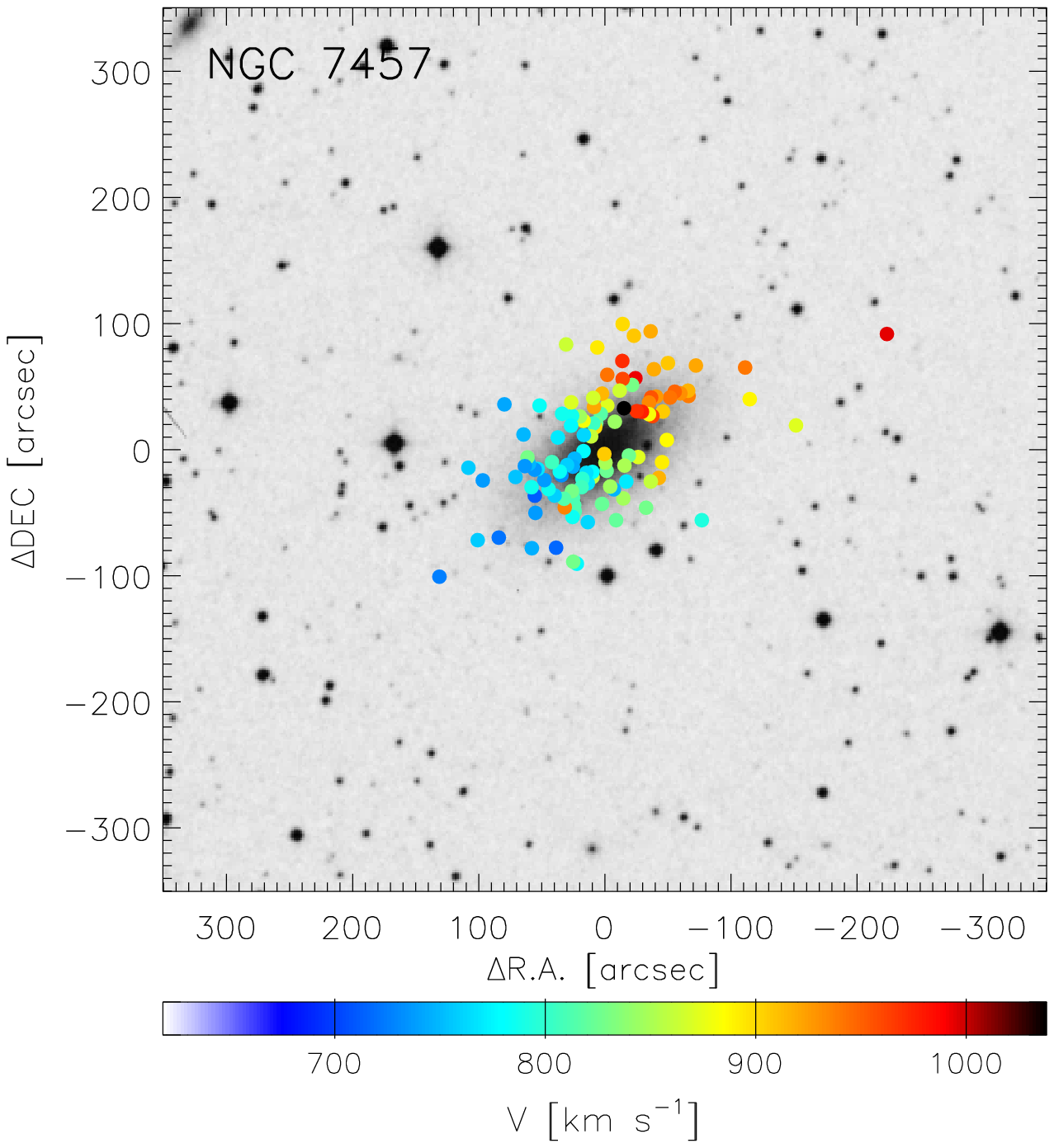}\\
\vspace{10pt}
\includegraphics[width=0.4\textwidth,height=0.4\textwidth]{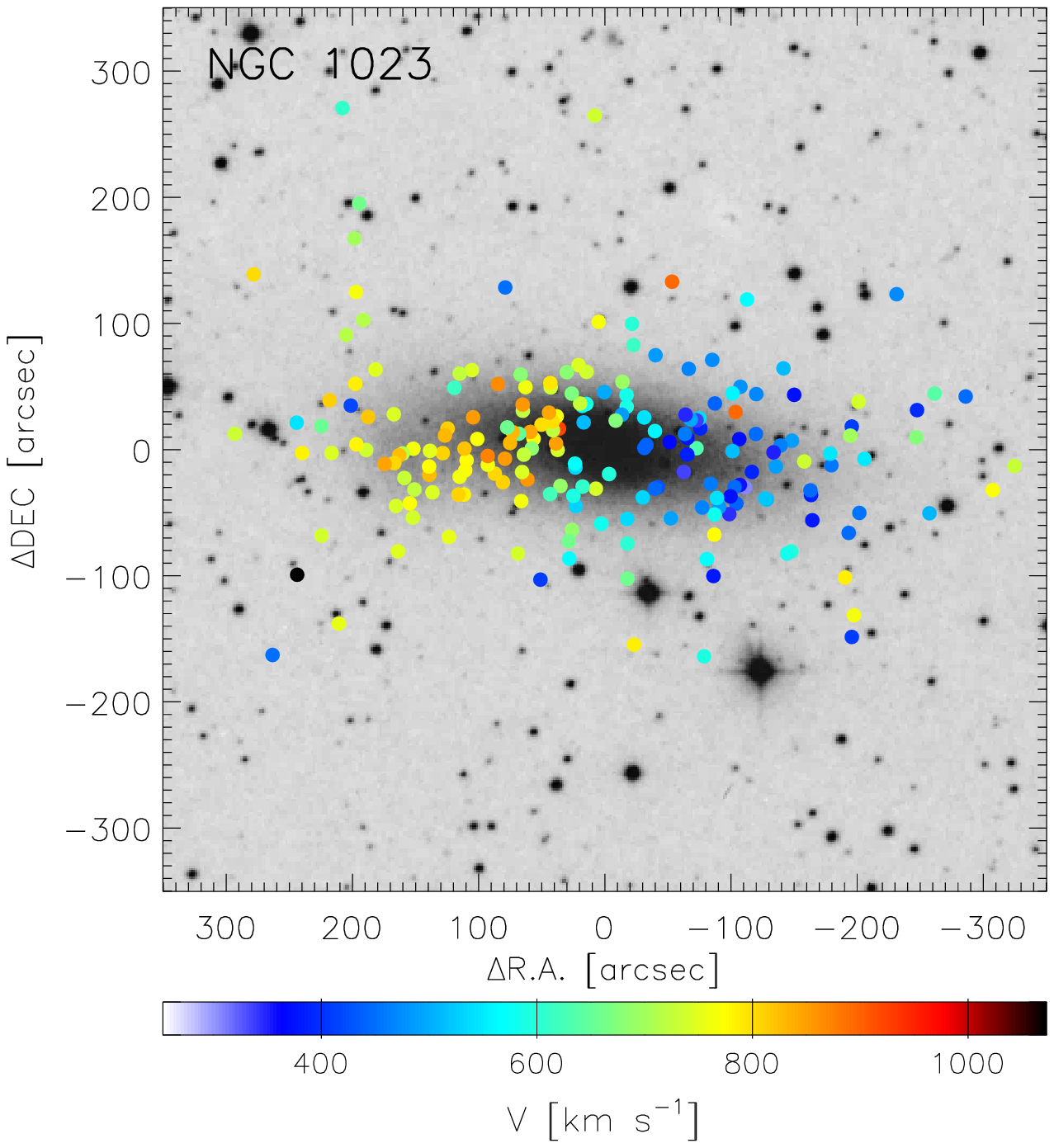} &  \includegraphics[width=0.4\textwidth,height=0.4\textwidth]{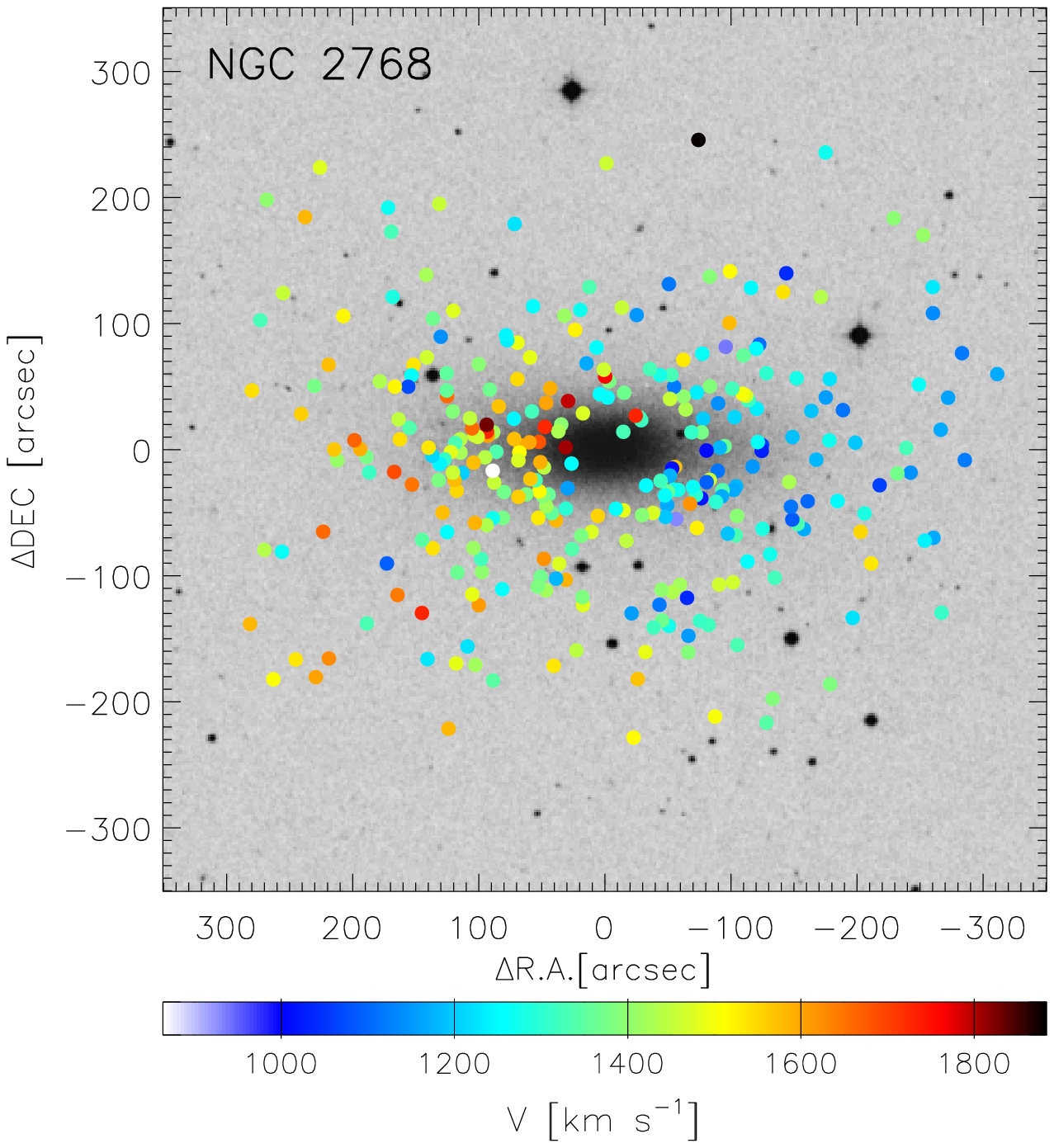}\\
\vspace{10pt}
\includegraphics[width=0.4\textwidth,height=0.4\textwidth]{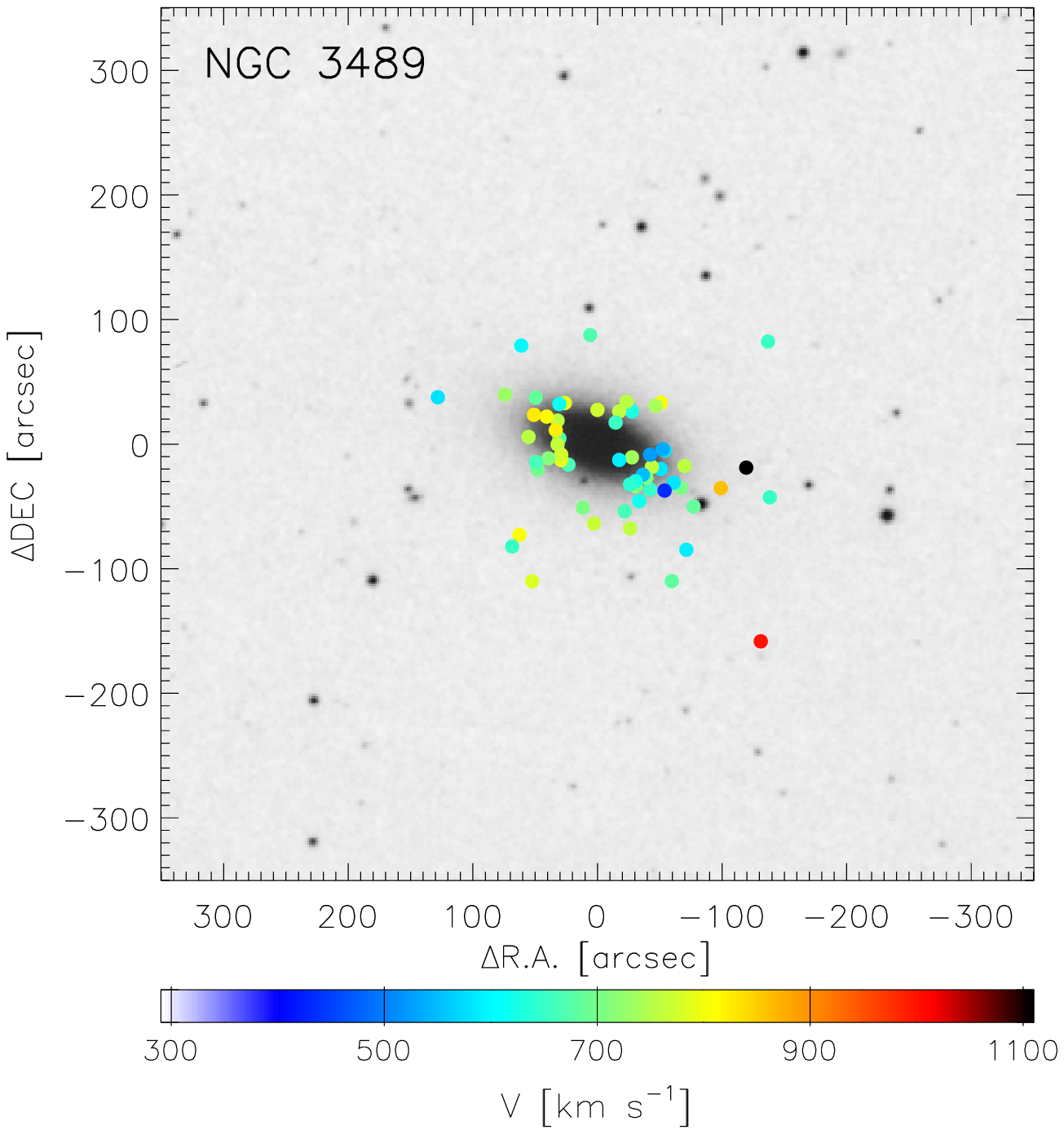} &  \includegraphics[width=0.4\textwidth,height=0.4\textwidth]{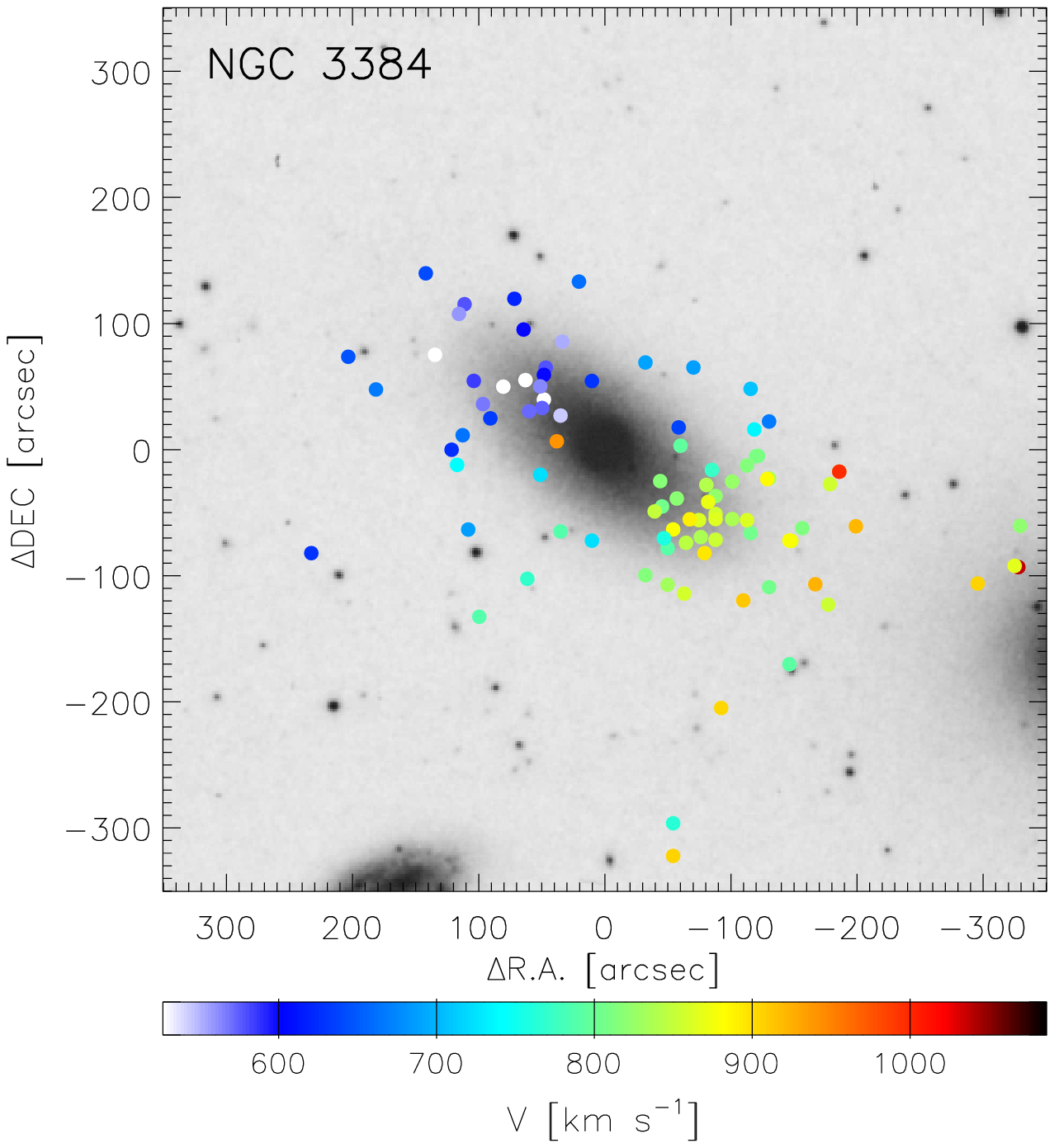}\\
\end{tabular}
\vspace{10pt}
\caption[PNe kinematics for the S0 sample.]{DSS images of the sample galaxies, with the
PNe positions marked with circles. The color of the circles represent
the measured radial velocity of the PNe, according to the velocity
scales below each panel. First row: field S0s, second row: S0s in
groups, third row: S0s in rich groups.  \label{fig:overvel}}
\end{figure*}

In this section we present the resulting stellar
kinematics.  Figure~\ref{fig:overvel} shows the PNe data superimposed
on the  Digitised Sky Survey (DSS) image of each galaxy in the sample.  The
incompleteness at small radii due to the bright galaxy continuum,
discussed in Section~\ref{sec:PNe surface density}, is very apparent.
However, these plots also show the large radii to which PNe can be
detected in these data, which underlines how nicely they complement
conventional kinematic data: the point where the continuum becomes too
faint to obtain good absorption-line data is just where PNe become
readily detectable, and they remain present in significant numbers out
to the very faint surface brightnesses illustrated in
Figure~\ref{fig:pneandstarlight}.  

It is apparent from Figure~\ref{fig:overvel} that most of these
galaxies are characterised by obvious rotation, although there is also
 a significant amount of random motion.  It also appears
  that the balance between ordered and random motion differs from
  galaxy to galaxy; for example NGC~7457 seems to show more coherent
  motions than NGC~3115. To quantify such differences in a preliminary
  kinematic analysis, we derive the rotation velocity and random
  motions assuming that each galaxy can be described as a
  single-component thin disk.  Following \citet{Ari}, a maximum
  likelihood analysis is employed to determine in a series of radial
  bins the mean streaming velocity and the two components of random
  motion in the tangential and radial directions, here assumed to be
  coupled together by the epicycle approximation.  To minimise distortion of the fit  by outlier PNe, the analysis
  iteratively rejects objects whose deviation from the mean velocity
  in each bin is higher than 2.1 $\sigma$; in no case are more than
  $7\%$ of the total number of PNe rejected by this process, offering
  some reassurance that such a disk model is not unreasonable.

\begin{figure*}[!t] 
\begin{tabular} {lr}
\includegraphics[width=0.45\textwidth]{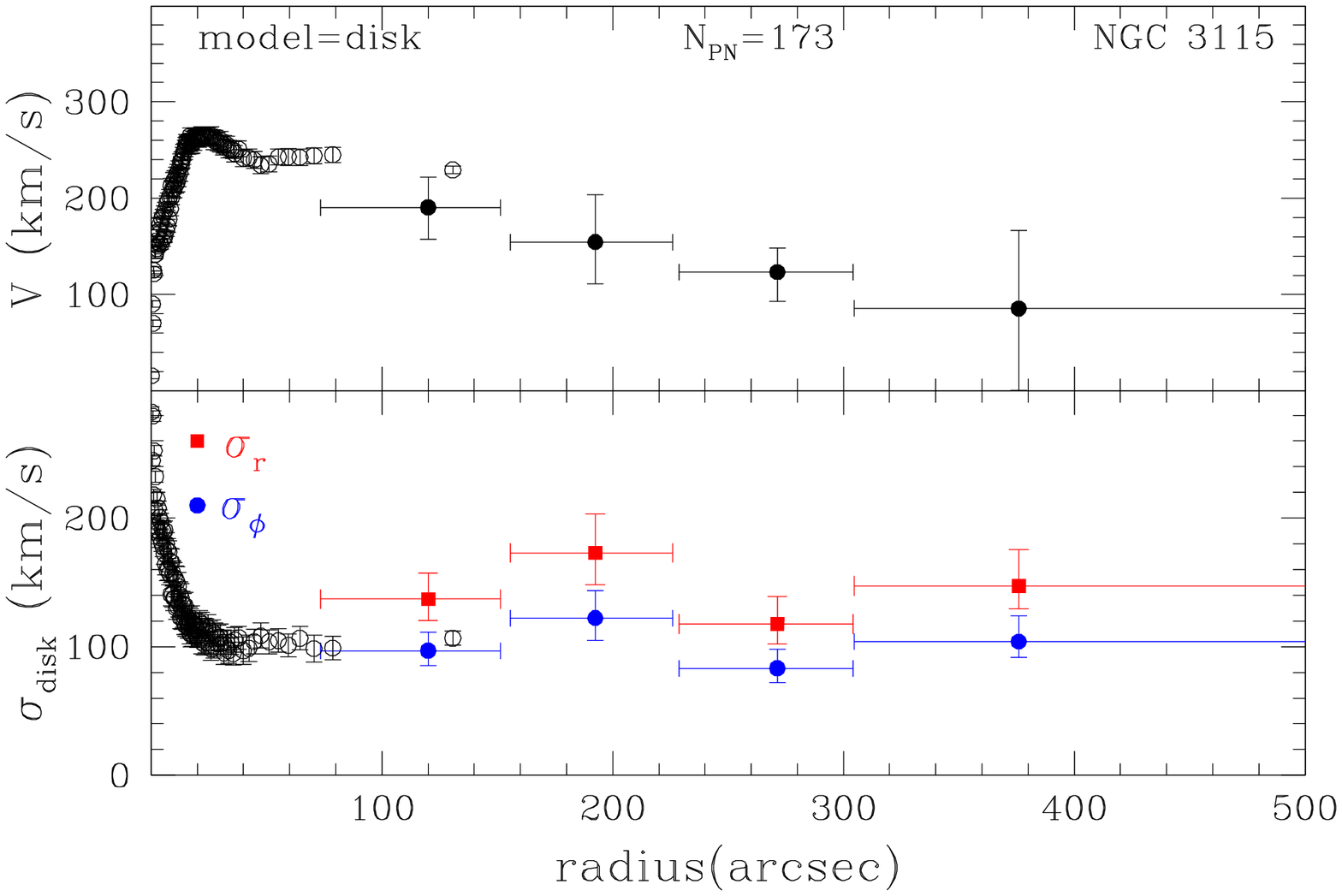} &  \includegraphics[width=0.45\textwidth]{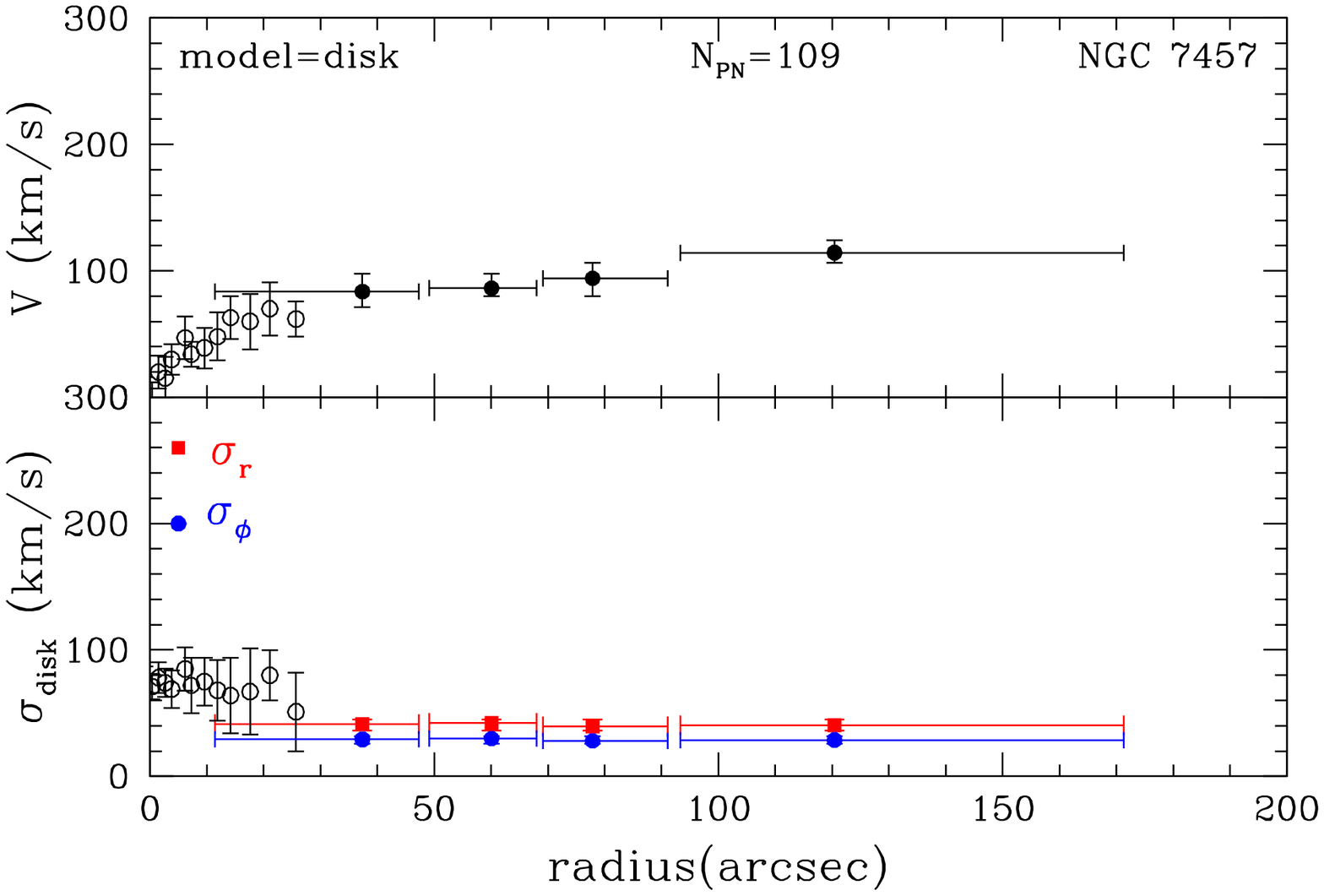}\\
\includegraphics[width=0.45\textwidth]{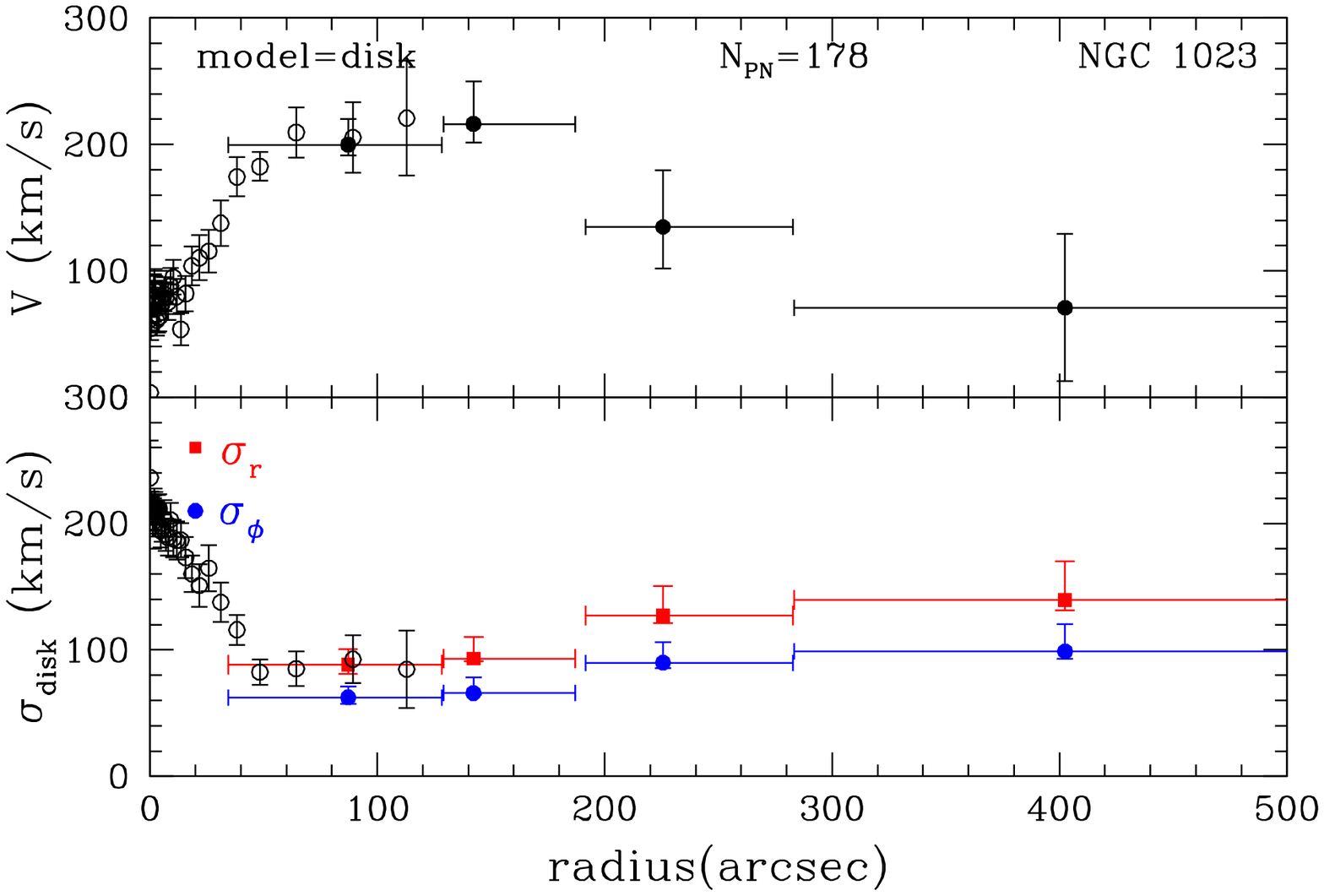} &  \includegraphics[width=0.45\textwidth]{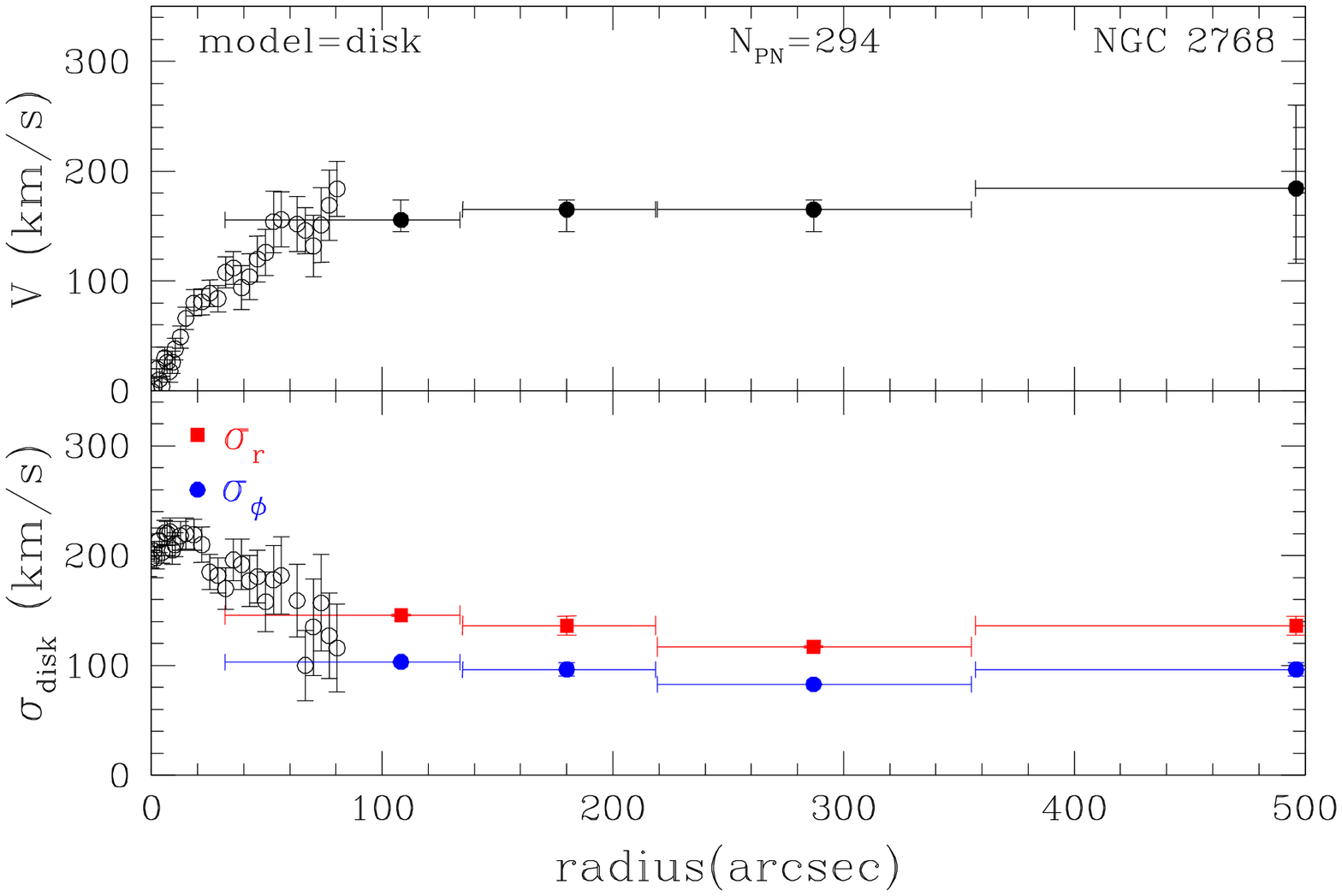}\\
\includegraphics[width=0.45\textwidth]{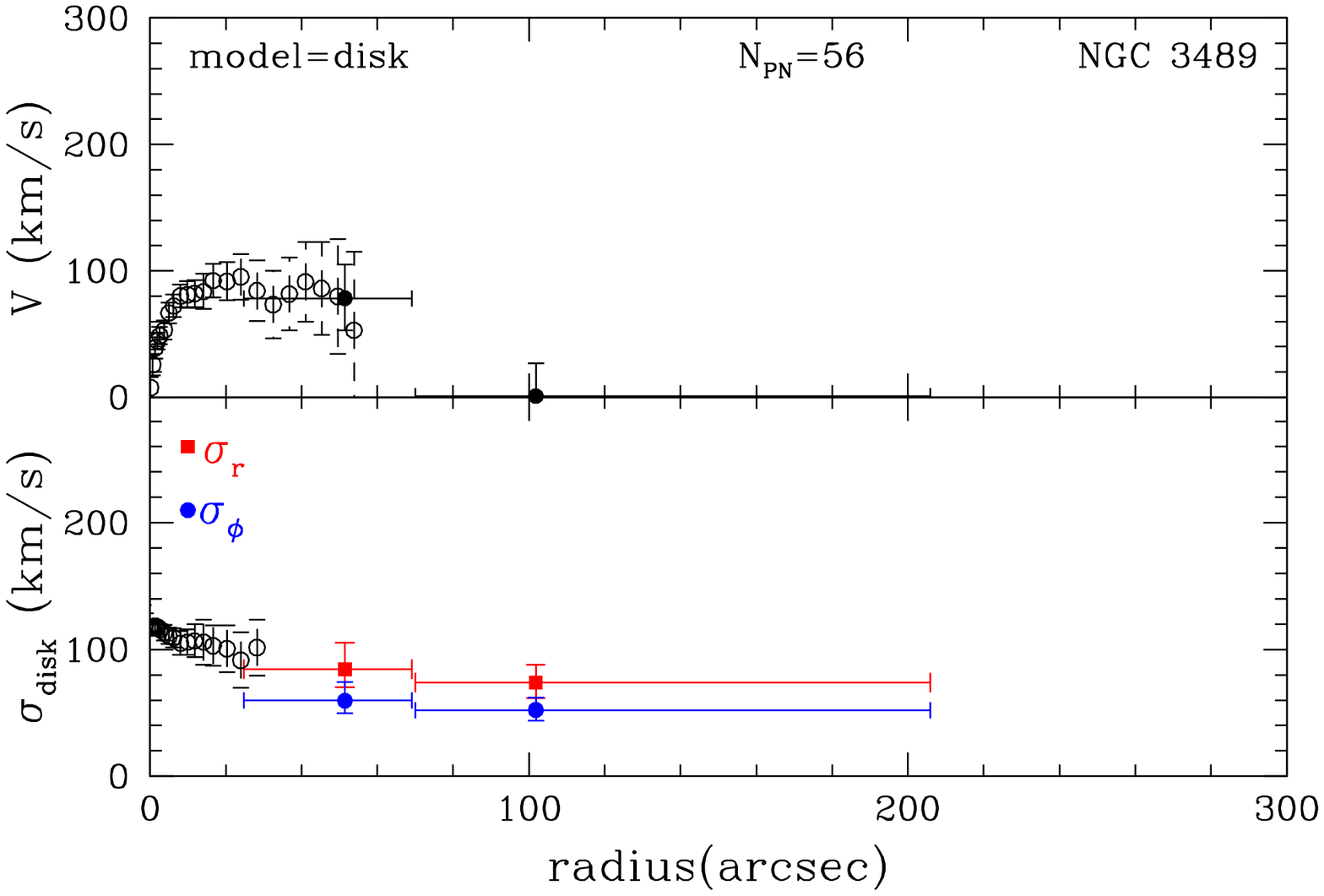} &  \includegraphics[width=0.45\textwidth]{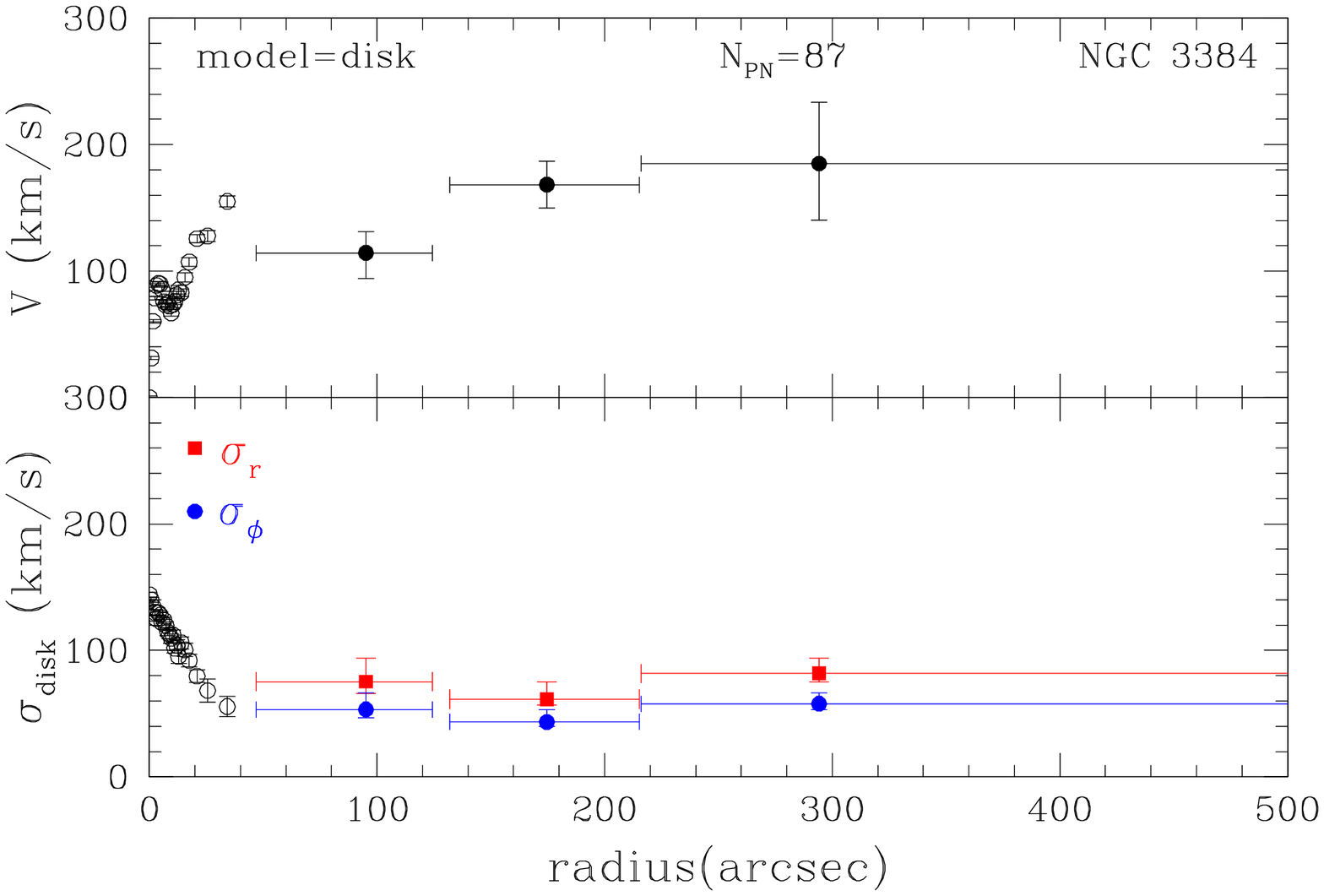}\\
\end{tabular}
\caption[Kinematics of the S0 sample assuming a disk-only model.]{
    Kinematics of the S0 sample derived using the disk model of
    \citet{Ari}. {\it First row}: field galaxies, {\it second row}:
    group members, {\it third row}: rich group members. The black
    circles show the average velocity in the bin (whose size is
    represented by the horizontal error bars). The red squares and
    blue circles show the velocity dispersion along the radial and
    azimuthal directions respectively.  The stellar velocities and
    velocity dispersions obtained from absorption line spectra
    \citep{Caon,Simien,Norris,Fisher97} are shown as empty black
    circles. The number of PNe left after the clipping performed in each likelihood fit is annotated at the top of the corresponding
    panel. \label{fig:diskvel}}
\end{figure*}

  As Figure~\ref{fig:diskvel} illustrates, this analysis confirms
  the impression that the balance between ordered and random motion
  varies significantly between galaxies.  It also offers one final
  very useful check on the reliability of PNe as stellar kinematic
  tracers, as we can compare the results to those from conventional
  absorption-line spectra. Figure~\ref{fig:diskvel} therefore also
  shows such absorption-line kinematics for the major axes of these
  galaxies, as obtained from various published sources
  \citep{Deb,Caon,Simien,Norris,Fisher97}.  Although the comparison
is not exact, these systems are sufficiently close to edge on that the
conventional major-axis velocity dispersion data should lie close to
$\sigma_\phi$ in the PNe data if both are measuring the same kinematic
population.  These plots underline the complementarity between
conventional absorption-line data and PNe kinematics, as the latter
take over just where the former run out.  It is also clear that there
is no inconsistency between these data sets, demonstrating once again
that they are probing the same kinematic population.

\section{Conclusions}
\label{sec:conclusion}

This paper has presented the data for the first systematic survey of S0
kinematics as a function of environment, using PNe to trace their
stellar velocities out to large radii.
In the process, we have presented a series of tests on the viability
of using PNe for such a study, demonstrating that they really are
representative of the underlying stellar population.  The PNe pass all
these tests: they show the same spatial distribution as the starlight;
their numbers are consistent with those found in comparable elliptical
galaxies where their use as kinematic tracers is well established;
and their kinematics join smoothly onto those derived from
conventional stellar data at smaller radii.  

As well as demonstrating that PNe are good tracers of the stellar
populations, and therefore that the catalogues presented in this paper
are a unique dataset to study the kinematics and dynamics of
lenticular galaxies, we have also shed some light on the nature
of S0 galaxies.  In particular, the similarity of their $\alpha$
parameters to those found in elliptical galaxies implies that, at
least at the stellar population level, these two classes of object are
very similar.  Unsurprisingly their kinematics are rather
different, being in most cases dominated by rotational motion.  There
is however significant variety in these kinematics, with rotation
velocities that vary in differing ways with radius and differing
balances between ordered and random motion.  The initial inspection of
these properties in Figure~\ref{fig:diskvel} does not reveal any
obvious trends with the richness of environment, suggesting that this
may not be a dominant factor in the choice of channel by which S0s
form.  However, \citet{Ari} already established that these simple
kinematic properties can be rather misleading due to
cross-contamination between disk and bulge kinematics, and that a
clear picture only emerges when the kinematic components are carefully
separated.  Indeed, by breaking S0 galaxies down into these
components, we hope to find subtler indications as to where
these systems lie in the overall story of galaxy morphology by, for
example, comparing their spheroidal components to ellipticals and to
the bulges of spirals.   We will present such an analysis in
a subsequent paper, to seek the definitive answers from the survey data
presented here.

\section*{Acknowledgments}

AC acknowledges the support from both ESO (from a  studentship in
2011 and a visitor program in 2012) and MPE (visitor program 2012).
LC acknowledges funding from the European Communityâ Seventh
Framework Programme (FP7/2007-2013/) under grant agreement No
229517. AJR was supported by National Science Foundation grant
AST-0909237. The PN.S team acknowledge the support of the UK and NL
WHT time allocation committees, and the support of the WHT staff
during several observing runs for the data acquisition on which this
paper is based. We thank the referee, W. J. Maciel, for the useful comments. 
We thank K. T. Smith for carefully proofreading this paper.
This research has made use of the 2MASS data archive,
the NASA/IPAC Extragalactic Database (NED), and of the ESO Science
Archive Facility.


 
\bibliography{PNS.DATA}
\bibliographystyle{aa}

\end{document}